\documentclass{aa}  
\usepackage{graphicx}
\usepackage{txfonts}
\begin{document}
\title{The connection between Gamma-ray bursts and Supernovae Ib/c}

\author{Elisabetta Bissaldi$^{\,1}$, Francesco Calura$^{\,2}$, 
Francesca Matteucci$^{\,2,3}$, Francesco Longo$^{\,1,4}$ and Guido Barbiellini$^{\,1,4}$}
\institute{
1 Dipartimento di Fisica, Universit\'a di Trieste, Via Valerio 2, 34127, Trieste, Italy \\
2 INAF - Osservatorio Astronomico di Trieste, via G. B. Tiepolo 11, 34131 Trieste, Italy \\
3 Dipartimento di Astronomia - Universit\'a di Trieste, Via G. B. Tiepolo 11, 34131 Trieste, Italy \\
4 INFN, Sezione di Trieste, Via Valerio 2, 34127, Trieste, Italy \\
}

\offprints{E. Bissaldi, currently at Max-Planck Institut f\"{u}r Extraterrestriche 
Physik, Giessenbachstra\ss{}e 1, 85748, Garching, Germany, \email{ebs@mpe.mpg.de} }

\date{Accepted}

\abstract
{It has been established that 
Gamma-Ray Bursts (GRBs) are connected to Supernovae (SNe) explosions of Type Ib/c.} 
{We intend to test whether the hypothesis of Type Ib/c SNe from different massive progenitors 
can reproduce the local GRB rate as well as the GRB rate as a function of redshift. 
We aim to predict the GRB rate at very high redshift under different assumptions about 
galaxy formation and star formation histories in galaxies.}
{We assume different star formation histories in 
galaxies of different morphological type: ellipticals, spirals and irregulars, which 
have been already tested in self-consistent galaxy models 
reproducing both chemical and photometrical properties of galaxies. 
We explore different 
hypotheses concerning the progenitors of Type Ib/c SNe: i) single massive stars ($M> 25 M_{\odot}$, 
Wolf-Rayet stars), ii) massive close binaries (12-20$M_{\odot}$) 
and iii) both Wolf-Rayet stars and massive binaries. 
We conclude that the mixed scenario 
(iii) is preferable to reproduce 
the local Type Ib/c SN rates in galaxies and we adopt this scenario for comparison with the GRB
rates.} 
{We find an excellent agreement between the observed GRB local rate and 
the predicted Type Ib/c SN rate in irregular galaxies, when a range for single Wolf-Rayet stars 
of 40-100$M_{\odot}$ is adopted. 
We also predict the cosmic Type Ib/c SN rate by taking into 
account all the galaxy types in an unitary volume of the Universe and we compare it with the 
observed cosmic GRB rate as a function of redshift.
By assuming the formation of spheroids at high redshift, we predict a cosmic Type Ib/c SN rate, 
which is 
always higher than the GRB rate, suggesting 
that only a small fraction (0.1-1$\%$) of Type Ib/c SNe become GRBs. 
In particular, we find 
a ratio between 
the cosmic GRB rate and the cosmic Type Ib/c rate in the range $10^{-2}-10^{-3}$, in agreement 
with previous estimates. Finally, due to the high star formation in spheroids at high redshift, which is our preferred scenario for galaxy formation, 
we predict more GRBs at high redshift than in the hierarchical scenario for galaxy formation, a prediction which awaits to be 
proven by future observations.}{}


\authorrunning{Bissaldi et al.}
\maketitle
\section{Introduction}\label{secIntro}

Gamma Ray Bursts (GRBs) are sudden and powerful flashes of gamma-ray radiation,
occurring random at a rate of $\sim$ 1 per day in the Universe.
The duration of GRBs at MeV energies ranges from $10^{-3}$ sec to about $10^{3}$ sec, 
with long bursts being characterized by  a duration $> 2$ seconds.   

During the last years, it has been established that
at least a large fraction of long-duration GRBs is directly
connected with the death of massive stars.
This scenario has got strong support from 
observations of supernova (SN) features
in the spectra of several of GRB afterglows.
Examples of the spectroscopic SN-GRB connection include
SN~1998bw/GRB~980425 (Galama et al. 1998), 
SN~2003dh/GRB~030329 (Stanek et al. 2003, Hjorth et al. 2003), SN~2003lw/GRB~031203 (Malesani et al. 2004) 
and SN~2006aj/GRB~060218 (Masetti et al. 2006, Campana et al. 2006, Modjaz et al. 2006). 

Originally, SNe were classified in to two main types (I, II) on the basis 
of the presence (Type II) or absence (Type I) of H lines in their spectra
(Zwicky 1938). At the present time we distinguish among five SN types on the 
basis of the spectra obtained at maximum light. 
Classical Type II SNe have prominent Balmer lines
exhibiting P-Cygni profiles and represent about 70\% of the exploding
stars in the Universe (Cappellaro et al. 1999). The classification can be further
subdivided in Types II-L and II-P, according to the shape of the light curve 
which can be linear (L) or have a plateau (P).Type IIn SNe have strong H lines in
emission. They can be distinguished from the classical Type II SNe by
the lack of absorption in their Balmer lines. \cite{CHU97} introduced
this designation to reflect the fact that these SNe undergo significant 
interaction with a ``dense wind'' produced by the SN progenitor
prior to explosion. Type Ia SNe are characterized by the lack of H lines
and by a strong absorption
observed at $\lambda \lambda$ 6347, 6371 \AA\, attributed 
to the P-Cyg profile of Si II. Type Ib SNe are identified by spectra
with no evident Balmer lines, weak or absent Si II lines and strong He I lines.
\cite{BER64} reported the first observation of this class of SNe, 
but the ``Ib'' designation was introduced later by \cite{ELI85}.
Type Ic SNe are characterized by
weak or absent H and He lines and no evident Si II. They show Ca II 
H\&K in absorption, the Ca II near-IR triplet with a P-Cygni profile,
and O I in absorption. 
The ``Ic'' class was introduced by \cite{WHE86}. 

Interestingly, the SN features observed in the afterglows of GRBs
resemble those of Type Ib/c SNe (Della Valle 2006; Woosley \&\ Heger 2006, 
and references therein).
In particular, modeling of the SN lightcurves reveals that
the SNe with GRB-connection have very large explosion
energy and mass production of $^{56}$Ni compared to normal
Type Ib/c SNe (Iwamoto et al. 1998, Nakamura et al. 2001, 
Deng et al. 2005, Mazzali et al. 2006b), except SN 2006aj
which requires an explosion energy that is comparable to
that of normal SNe Ib/c (Mazzali et al. 2006a). These facts
have motivated people to invent the term ``hypernovae'' for
this special and much more powerful class of SNe (Iwamoto
et al. 1998; see also Paczy$\grave{\rm n}$ski 1998).

These phenomenological results greatly support the ``collapsar'' model 
(MacFayden \& Woosley 1999, 
Zhang, Woosley \& MacFayden 2003) 
in which a Wolf-Rayet progenitor undergoes core collapse, 
producing a rapidly rotating black hole surrounded by an 
accretion disk which injects energy into the system and 
thus acts as a ``central engine''. The energy extracted
from this system supports a quasi-spherical Type Ib/c SN
explosion and drives collimated jets through the stellar rotation
axis which produce the prompt gamma-ray and afterglow emission
(e.g., Zhang \&\ M\'esz\'aros 2004).

For what said before, it is clear that a comparison between 
the Type Ib/c SN rates in galaxies and the GRB rate is very 
important in order to test if the SN-GRB connection is in 
agreement with the actual knowledge about SN progenitors, 
stellar and galactic evolution.

Galaxies of different morphological type evolve in a different 
way with the main driver of their evolution being the star 
formation history. Recently, Calura \&\ Matteucci (2006) calculated 
the rates of Type II, Ib and Ic SNe in galaxies of different morphological 
type (ellipticals, spirals and irreguòlars), by means of self consistent 
galaxy evolution models, reproducing both the chemical and spectro-photometric 
properties of galaxies.

In this paper, we present a study of models which accurately reproduce 
the local Type Ib/c SN rate in galaxies, by following the approach of
Calura \&\ Matteucci (2006). We assume different progenitors for
the Type Ib/c SNe such as single Wolf-Rayet (WR) stars and massive stars 
($12-20 M_{\odot}$) in close binary systems, 
as well as several star formation 
histories typical of galaxies of different morphological type.

The assumed range for the binary progenitors of Type Ib/c SNe are based 
on suggestions by Baron (1992), Nomoto et al. (1996) and on the 
recent modelling of the 
proto-type Ic supernova SN1994 I (Sauer et al. 2006), which suggests 
that the 
upper mass limit for these binary stars should not be larger than 
$20 M_{\odot}$. For what concerns 
the single WR stars as Type Ib/c progenitors, we need to assume a lower 
mass limit for the formation  of WR stars but this limit in uncertain 
and strongly depends on the details of stellar evolution models such as rotation and 
assumed mass loss, which in turn depends on the initial stellar chemical 
composition (see Meynet \&\ Maeder (2005) and Yoon, Langer \& Norman (2006) for a recent study 
on the single WR stars and GRBs). Therefore, here we will explore several cases where the lower mass for the 
formation of a single WR star varies from 25 up 90 $M_{\odot}$. However,
it is important to recall that most stars with $M > 25 M_{\odot}$ probably 
produce black holes, either in a prompt collapse or by fallback; in neither 
case one would expect a normal Type Ib/c SN,  since probably most of the mass 
in the Si shell, which 
is turned into Ni during the explosion, would fall into the black hole. 
So, it is still unproven that 
massive SNe forming black holes would appear as typical SNe Ib/c. 
If, instead, such massive stars are in interacting binary systems, their 
final core evolution is drastically changed and they are expected to  
produce neutron stars rather than black holes (Brown et al. 2001) and hence 
possibly become SNe Ib/c. All these facts, seem to suggest that the binary 
channel for the formation of SNe Ib/c is favored over the channel of 
single very massive stars.

We then derive the SN rates and compare them  with 
the observed local GRB rates, thus obtaining an estimate of the local GRB/SN 
ratio.
Furthermore, we move on cosmological scales by computing the cosmic  
Type Ib/c SN rates and, on the basis of that, we chose the best model 
for the GRB distribution as a function of redshift and therefore we 
predict the expected number of GRBs at high redshift.

The paper is organized as follows: in Section 2 we present the galactic 
models and the theoretical Type Ib/c SN rates for galaxies of different 
morphological type. In Section 3 we discuss the local GRB rate and 
compare it with the theoretical predictions. In Section 4 we describe 
how we compute the cosmic Type Ib/c SN rates and we extract a best model 
for the distribution of GRB as a function of redshift. Finally in Section 
5 some conclusions are drawn.

\subsubsection{Evolution Models for Galaxies}\label{SubSecEvo}
In order to compute the SN Ib/c rates and compare them with the observed 
GRB rate we adopted self-consistent galaxy evolutionary  models already 
reproducing the majority of the properties of galaxies of different 
morphological type. In particular, these models are tuned to reproduce 
the present time SN rates and the chemical abundances in stars and gas. 
Moreover, the predicted star formation and chemical enrichment histories, 
when adopted in spectro-photometric models, give good agreement with
photometric properties of galaxies.

In order to better understand the adopted galaxy models, a schematic outline
of their basic equations and physical processes is provided (Calura \& Matteucci 2003).

Let $G_{i}(t)$ be the normalized fractional mass of gas within a galaxy
in the form of the element $i$:
\begin{equation}\label{eqa1}
G_i(t)\;=\;\frac{M_g(t)\,X_i(t)}{M_{tot}} \, ,
\end{equation}
where $M_{tot}$ and $M_g(t)$ are the total galaxy mass 
and the mass of gas at time $t$, respectively.
\begin{equation}\label{eqa2}
X_{i}(t)\;=\;\frac{G_{i}(t)}{G(t)}
\end{equation}
represents the abundance in mass of the element $i$,
the summation over all elements in the gas mixture being equal 
to unity.
Thus, the quantity 
\begin{equation}\label{eqa3}
G(t)= \frac{M_{g}(t)}{M_{tot}}
\end{equation}\label{eqa4}
is the total fractional mass of gas present in the galaxy at time $t$.

The temporal evolution of $G_{i}(t)$ is described by the basic equation:
\begin{equation}\label{eqa5}
\dot{G_{i}}=-\psi(t)X_{i}(t) + R_{i}(t) + \left(\dot{G_{i}}\right)_{inf} - 
\left(\dot{G_{i}}\right)_{out}\, .
\end{equation} 
Here, $\psi(t)$ is the star formation rate (SFR), 
namely the fractional amount of gas turning into stars per unit time.
$R_{i}(t)$ represents the returned fraction of matter in the 
form of an element $i$ that the stars eject into the interstellar medium (ISM) through 
stellar winds and SN explosions.
This term contains all the prescriptions regarding the stellar yields and
the SN progenitor models.
The two terms $\left(\dot{G_{i}}\right)_{inf}$ and $\left(\dot{G_{i}}\right)_{out}$ 
account for the infalling external gas from the intergalactic 
medium and for the outflow, occurring by means of SN driven galactic 
winds, respectively. 

Here we are interested in computing the SN rates in the framework of these self-consistent galaxy models.
The two key ingredients for computing the SN rates are
the SFR and the stellar
initial mass function (IMF). The first is generally expressed
as a function of time only, while the second, which describes
the stellar mass distribution at birth, is likely to be universal
(Kroupa 2002) and not vary as
a function of time (Chiappini et al. 2000).

\subsubsection{Star Formation Rates}\label{SubSecSFR}
The main feature characterizing a particular morphological galactic type is
represented by the prescription adopted for the star formation history,
summarized in the SFR expression. \\
In the case of elliptical 
galaxies, the SFR $\psi(t)$ (in $Gyr^{-1}$) has a simple form and is given by:
\begin{eqnarray}\label{eqa6}
\psi(t) = \nu\, G(t) 
\end{eqnarray}
The quantity $\nu$ is the efficiency of star formation, namely the inverse of
the typical time scale for star formation. 

In the case of ellipticals, $\nu$ is
assumed to drop to zero at the onset of a galactic wind, which develops as the
thermal energy of the gas heated by supernova explosions exceeds the binding
energy of the gas (Arimoto \& Yoshii 1987, Matteucci \& Tornamb\'e 1987). 
This quantity is strongly
influenced by assumptions concerning the presence and distribution of dark
matter (Matteucci 1992); for the model adopted here a diffuse 
($R_e/R_d$=0.1, where
$R_e$ is the effective radius of the galaxy and $R_d$ is the radius 
of the dark matter core) but 
massive ($M_{dark}/M_{Lum}=10$) dark halo has 
been assumed.

In the case of irregular galaxies, a continuous star formation 
rate is assumed as in eq. \ref{eqa6}, but
characterized by a lower efficiency than the one adopted for ellipticals.

In the case of spiral galaxies, the SFR expression is slightly more complex, 
since it takes into account the feedback between SNe and ISM by means of a dependence 
upon the total surface mass density (see Chiosi 1980, Chiappini et al. 1997):
\begin{eqnarray}\label{eqa7}
\psi(r,t) \; = \; \nu \, \left[\frac{\sigma(r,t)}{\sigma(r_{\odot},t)}\right]^{2(k-1)}
 \left[\frac{\sigma(r,t_{Gal})}{\sigma(r,t)}\right]^{k-1}
 G^k (r,t)
\end{eqnarray}
where $\nu$ is the SF efficiency, $\sigma(r,t)$ is the total 
(gas + stars) surface mass density at a radius r and time t, and 
$\sigma(r_{\odot},t)$ is the total surface mass density in the 
solar region. 
For the gas density exponent $k$, a value of 1.5 has been assumed by \cite{CHI97}, in
order to ensure a good fit to the observational constraints at 
the solar vicinity and in agreement with the estimates by \cite{KEN98}.

\begin{figure*}
\centering
\includegraphics[width=0.95\columnwidth,bb=20 144 580 700,clip]{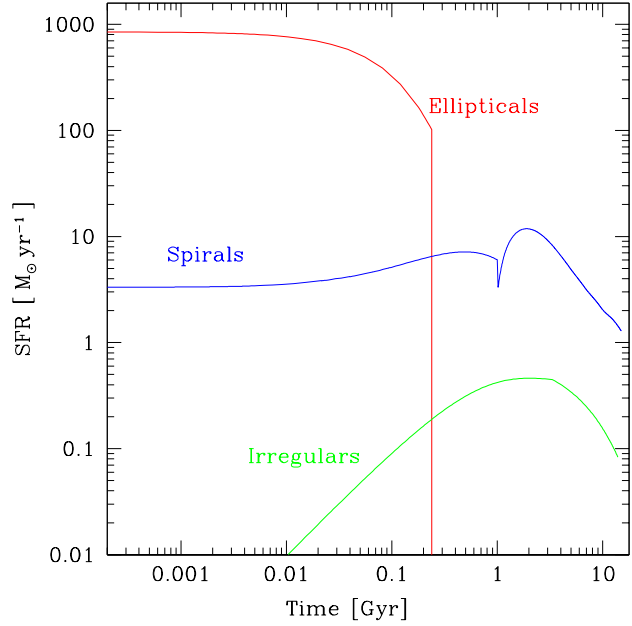}
\caption{Star formation rates, expressed in $M_{\odot}yr^{-1}$,  as functions of time
calculated according to equations \ref{eqa6} and \ref{eqa7} for elliptical 
({\it upper curve}), spiral 
({\it median curve}) and irregular ({\it lower curve}) galaxy models.}
\label{figP1}
\end{figure*}


The three different star formation rates for elliptical, irregular 
and spiral galaxy models as functions of time are shown in Figure \ref{figP1}. 
The SFR of the elliptical model (corresponding to a galaxy with luminous mass of $10^{11}M_{\odot}$)
is characterized by very high SFR values 
(from 100 to 1000 M$_{\odot}$ yr$^{-1}$) 
and by a starburst lasting $\sim$ 0.2 Gyr, which traces
the rapid collapse of a homogeneous sphere of primordial gas 
where star formation is taking place at the same time as the collapse proceeds.
Star formation halts as the energy of the ISM,
heated by stellar winds and SN explosions,
balances the binding energy of the gas. At this time a galactic wind
occurs, sweeping away almost all the residual gas.

The SFR of the irregular model is continuous and is 
characterized by a low SF efficiency.
In this scenario, irregulars are assumed to assemble from merging
of protogalactic small clouds of primordial chemical composition, until a mass
of $\sim$ 6 $\times$ 10$^{9}$ M$_{\odot}$ is accumulated,
therefore producing stars at a lower rate than spiral  and elliptical
galaxies.

The predicted SFR for spirals is characterized by two peaks due to 
two infall episodes present in the model of Chiappini et al. (1997).
During the first episode, the halo forms and the gas shed by the halo
rapidly gathers in the center, yielding the formation of the bulge.
During the second episode, a slower infall of external gas forms the disk,
with the gas accumulating faster in the inner than in the outer region
(``inside-out'' scenario; Matteucci \&\ Fran\c{c}ois 1989). The process
of disk formation is much longer than the halo and bulge formation, with
timescales varying from $\sim$ 2 Gyr in the inner disk to $\sim$ 7 Gyr in the solar
region and up to 15--20 Gyr in the outermost disk. 
The two infall episodes are clearly visible in Figure \ref{figP1}.
The second one is dominated by the contribution of
the disk. 
Here, the SFR for the MW represents an average
between 2 and 20 kpc from the galactic center, and it
is assumed as representative for an average spiral galaxy.

\subsubsection{Initial Mass Function}\label{SubSecIMF}
The most widely used functional form for the IMF is
an extension of that proposed by Salpeter (1955) 
to the whole stellar mass range:
\begin{eqnarray}\label{eqSalp}
\phi_{Salpeter} (M) = {\mathcal A}_{Salpeter} \; M^{-\,(1+x)},
\end{eqnarray}
where $x\,=\,1.35$, ${\mathcal A}_{Salpeter}\,\simeq\,0.17$ is the normalization constant derived from:
\begin{eqnarray}\label{eqnorm}
\int_{\,M_{low}}^{\,M_{up}}M\,\phi_{Salpeter}(M)\,dM = 1 \, .
\end{eqnarray}
The IMF in eq. (7) is by number with $M_{low}\,=\,0.1\,M_{\odot}$ and 
$M_{low}\,=\,100\,M_{\odot}$.
We are adopting this IMF for ellipticals and irregulars, where there is not a direct 
measure of the IMF. On the other hand,   
multi-slope expressions of the IMF give a better description
of the luminosity function of the main sequence stars in the
solar neighbourhood, as proposed by 
 \cite{TIN80} and Scalo (1986, 1998).
A simplified two-slope approximation to the actual Scalo (1986) IMF, 
that we adopt for the Milky Way and spirals, is given by:
\begin{eqnarray}\label{eqSca86}
\phi_{Scalo86}(M) = {\mathcal A}_{Scalo86} \; M^{-\,(1+x)},
\end{eqnarray}
where
\begin{eqnarray*}
x \,=\, 1.35 \quad {\mathcal A}_{Scalo86}\; \simeq \; 0.19 \quad if \; \;M\, \leq\, 2\, M_{\odot} \\
x \,=\, 1.7 \; \quad {\mathcal A}_{Scalo86}\; \simeq \; 0.24 \quad if \; \;M\, >\, 2\, M_{\odot} \, .
\end{eqnarray*}
Also for this IMF holds the normalization condition expressed in eq. (8) for the
mass range 0.1--100 $M_{\odot}$.
Finally, one can consider the so-called
``top-heavy'' IMF, which was suggested by observational
and theoretical arguments (e.g. Larson 1998 and references
therein) and which assumes that the mass scale of the IMF was 
higher at earlier stages of the Universe, i.e. at high redshift:
\begin{eqnarray}\label{eqTop}
\phi_{t-h} (M) = {\mathcal A}_{t-h} \; M^{-\,(1+x)},
\end{eqnarray}
where $x\,=\,1.0$ and ${\mathcal A}_{t-h}\,\simeq\,0.14$.
However, this kind of IMF does not produce satisfactory
results for the solar neighbourhood and it is better to
assume a constant IMF (Chiappini et al. 2000).

\subsection{SNR$_{\mathbf {Ib/c}}$ models}\label{SubSecSNR}
Generally, Type Ib/c SNe
occur in the vicinity of star forming regions and usual
candidates for their progenitors are represented
by massive WR stars (e.g. Hamuy 2003). However, the limiting mass 
for the formation of a WR star depends on the assumptions made 
in stellar models such as mass loss rate and rotation as recently 
shown by Meynet \&\ Maeder (2005). Their results suggest that the 
limiting mass for forming a WR star increases with decreasing metallicity
because of the growth of the mass loss rate with metallicity.
For the sake of simplicity we will start here by assuming 
that all massive stars with initial mass 
$M\,\gtrsim\,25\,M_{\odot}$ are becoming WR stars and explode as
type Ib/c SNe. Then we will test also larger values for this mass limit.

The type Ib/c SN rate  is then expressed as:
\begin{eqnarray}\label{eqModI}
{\rm SNR_{Ib/c}^{Model\,I}}(t) & = & \int_{\,25}^{\,100} \phi(M)\,\psi(t-\tau_M)\,dM \nonumber \\
                 & \simeq & \psi(t) \int_{\,25}^{\,100} \phi(M)\,dM\, ,
\end{eqnarray}
where $\tau_{M}$ is the stellar lifetime.
Hereafter, equation \ref{eqModI} will be referred to as Model I.

However, some observational evidence seems to be against the WR 
progenitor hypothesis. In particular, Type Ib/c SNe can show broader light 
curves than 
expected assuming  a massive star collapse origin.
Other problems are represented
by the apparent paucity of WR stars, unable to account
for the rates observed in local galaxies (Muller et al. 1992),
along with the fact that some events have been observed
relatively far from active star forming regions (Filippenko \&\ Sargent
1991). More promising candidates, which could account
for these evidences, are represented by less massive stars,
with typical masses of 12--20 M$_{\odot}$ (e.g. Baron 1992), 
in close binary systems. In this case, the
loss of their envelope occurs by means of Roche-Lobe overflows.
In this scenario (Model II), the SN rate can be expressed as:
$$
{\rm SNR_{Ib/c}^{Model\,II}}(t)  = F \int_{\,12}^{\,20} \phi(M)\,\psi(t-\tau_M)\,dM 
$$
\begin{equation}\label{eqModII}
\simeq F \, \psi(t) \int_{\,12}^{\,20} \phi(M)\,dM
\end{equation}
where $F$ is a parameter and represents the fraction of massive binary 
systems of that particular kind giving rise to Type Ib/c SNe.
This parameter is chosen to be equal to
0.15 (Calura \& Matteucci, 2006). This choice is 
motivated by the facts that first, in
any galaxy, half of the massive stars are in binary systems,
and second, that the fraction of massive stars
in close binary systems (namely those which can give rise to SNe Ib/c) 
 is reasonably $\sim$ 30\%, i.e. similar to
the close binary frequency predicted for low mass systems 
(Jeffries \& Maxted 2005). 
Therefore the estimated value for this parameter is given by:
\begin{eqnarray}\label{eqalfa}
F\,\simeq\,0.5\,\cdot\,0.3\,\simeq\,0.15\,. 
\end{eqnarray}

\begin{figure*}
\centering
\includegraphics[width=0.95\columnwidth,bb=28 144 580 700,clip]{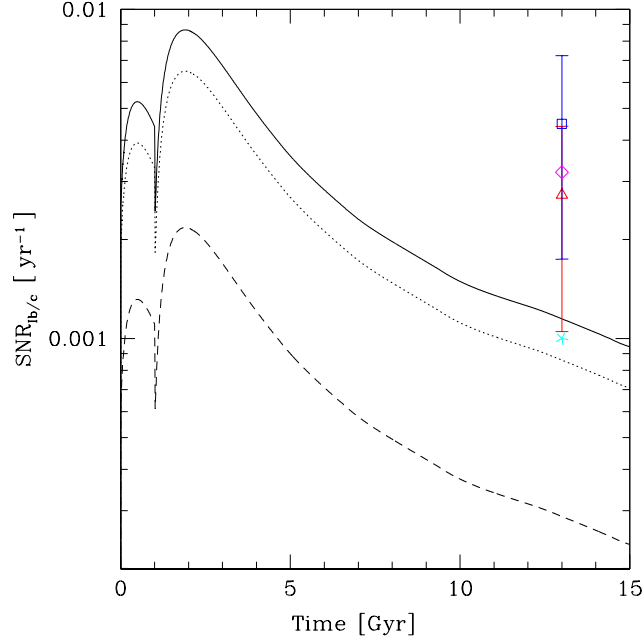}
\caption{SN rate for a spiral galaxy as a function of time
calculated according to Model I (equation \ref{eqModI}, {\it dotted line});
Model II (equation \ref{eqModII}, {\it dashed line}); 
Model III (equation \ref{eqModIII}, {\it solid line}).
Data from Cappellaro et al. (1999)
({\it open triangle}), Mannucci et al. (2005) 
({\it open square}), Della Valle (2005) ({\it open rhombus})
and Podsiadlowski et al. (2004) ({\it star}).A Scalo 1986 IMF is assumed.}
\label{figP2}
\end{figure*}

This is probably a lower limit for the 
fraction of binary systems giving rise to SNe Ib/c, as suggested 
in a paper by Podsiadlowski et al. (1992), where they concluded that 
15-30\% of all massive stars ($M > 8 M_{\odot}$),
can produce SNe Ib.
Moreover, a recent estimate by Kobulnicky et al. (2006) seems to 
favor the higher estimate. However, our choice of the binary fraction is 
also tuned to reproduce the observed SN Ib/c rates in galaxies, as shown by 
Calura \& Matteucci (2006), when also single WR stars are taken into account.
Therefore, taking both scenarios into account, a
cumulative SN rate is obtained (Model III):
$$
{\rm SNR_{Ib/c}^{Model\,III}}(t) = {\rm SNR_{Ib/c}^{Model\,I}}(t)
+\,{\rm SNR_{Ib/c}^{Model\,II}}(t) \\
$$
\begin{equation}\label{eqModIII}
\simeq  \displaystyle \psi (t) \left( \; F { \int_{\,12}^{\,20} \Phi (M) \, dM} 
+ { \int_{\,25}^{\,100} \Phi (M) \, dM} \right).
\end{equation}
In this case, the global fraction of Type Ib/c SN progenitors is similar to the above fractions of massive binaries by Podsiadlowski et al. (1992).
The three SN rates (Models I, II and III) as functions of time 
in the local Universe, i.e. for a spiral 
galaxy like the Milky Way,
are plotted in Figure \ref{figP2}. A Scalo (1986) IMF (eq. 9) is assumed. 
The two main infall episodes, which are characteristic 
for the halo and for the disk formation, respectively,  are clearly visible.


Observational rates measured by different authors are also 
indicated at present time ($t\,\sim\,13.47$ Gyr).
The SN rates are usually expressed in supernova units, SNus,
(1 SNu = 1 SN per 100 yr per 10$^{10}$ L$_{{\rm B}\,\odot}$;
1 SNuK = 1 SN per 100 yr per 10$^{10}$ L$_{{\rm K}\,\odot}$).
The rate of Type Ib/c SN exploding in spiral galaxies, i.e. type Sbc/Sd,
was measured by \cite{CAP99} obtaining:
\begin{eqnarray}
{\rm SNR_{Ib/c}}(\,t\sim\,13.47\,{\rm Gyr}\,)\,=\,\left(0.14\,\pm\,0.07\right)\;{\rm SNu} \, . \nonumber 
\end{eqnarray}
For the same rate, \cite{MAN05} calculated a value: 
\begin{eqnarray}
{\rm SNR_{Ib/c}}(\,t\sim\,13.47\,{\rm Gyr}\,)\,=\,\left(0.067\,^{\,+0.041}_{\,- 0.032}\right)\;{\rm SNuK}\,. \nonumber 
\end{eqnarray}
In order to compare these rates with the SN rate models,
which are calculated in units of SNe yr$^{-1}$,
one needs to multiply them by the correct galactic luminosities. 
For spiral galaxies, the Milky Way B-band luminosity
L$_{B}$ = (1.8 $\pm$ 0.3) $\times$ 10$^{10}$ L$_{B\,\odot}$
(van der Kruit 1986) and K-band luminosity 
L$_{K}\,\simeq\,6.7\,\times\,10^{10}$ L$_{K\,\odot}$
(Kent et al. 1991) are assumed. 

{\begin{figure*}
\centering
\includegraphics[height=19pc,width=19pc]{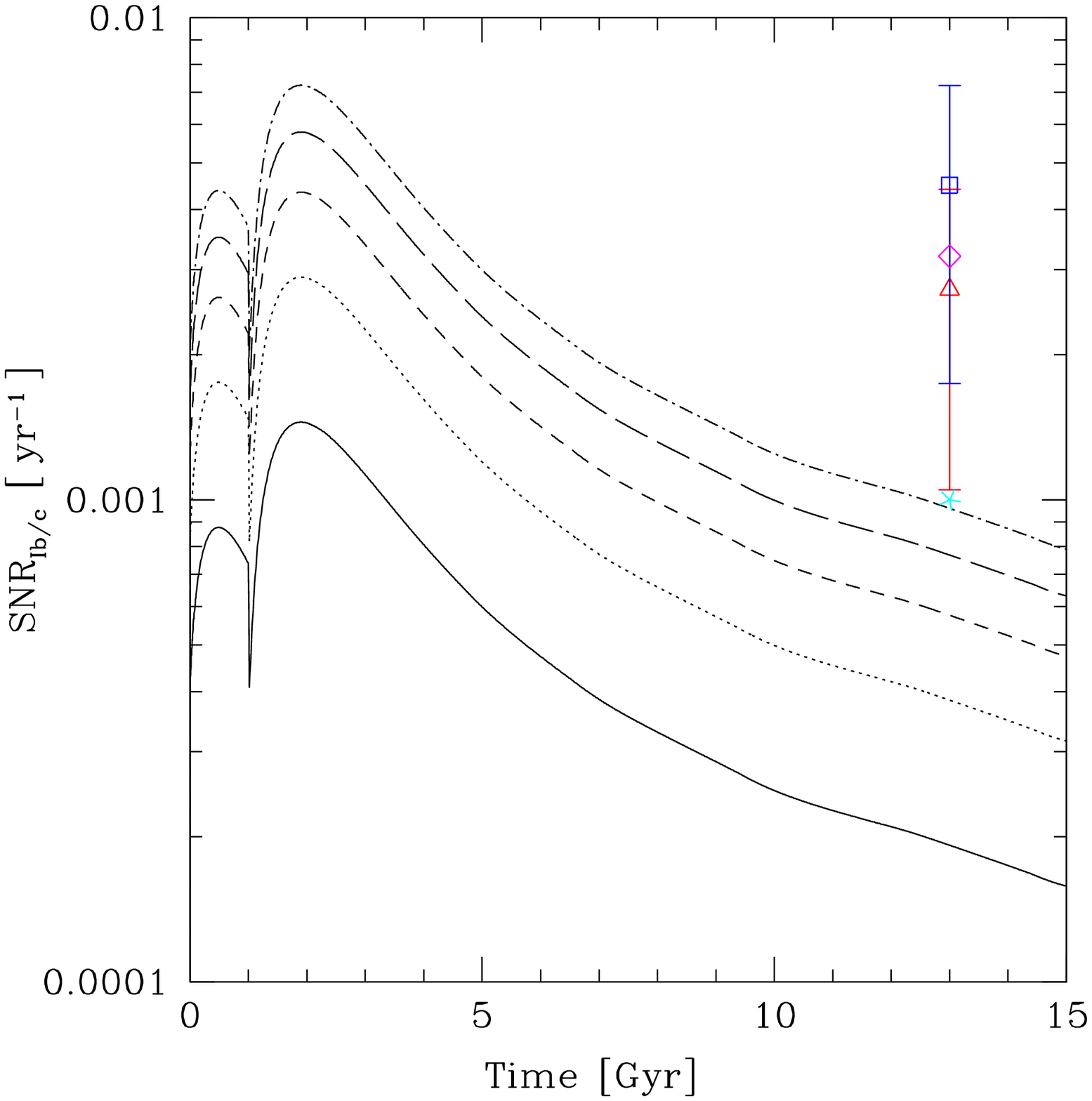}    
\includegraphics[height=19pc,width=19pc]{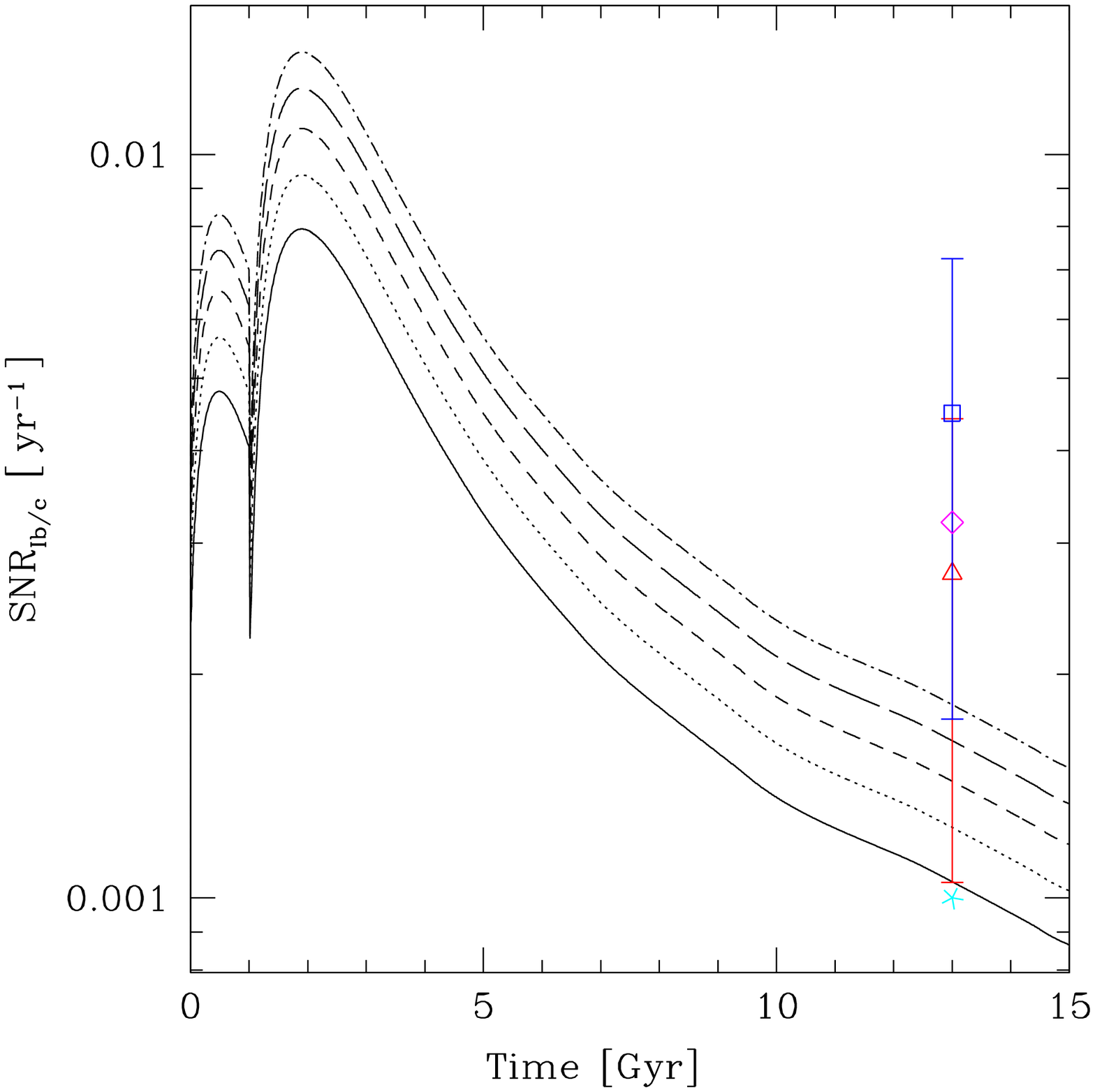}
\caption{SN rate for a spiral
galaxy as a function of time
calculated according to Model II (equation 12, {\it Left Panel})
and Model III (equation 14, {\it Right Panel}). Different values of the
parameter $F$ are considered: 0.1 ({\it
solid line}); 0.2 ({\it dotted line}); 0.3 ({\it
short-dashed line}); 0.4 ({\it long-dashed line});
0.5 ({\it dot-dashed-line}); Data from Cappellaro et al. (1999)
({\it open triangle}), Mannucci et al. (2005) 
({\it open square}), Della Valle (2005) ({\it open rhombus})
and Podsiadlowski et al. (2004) ({\it star})A Scalo 1986 IMF is assumed in both cases.}
\label{figP3}
\end{figure*}

Furthermore, the validity of the choice for the
adopted massive binary star parameter, namely $F$ = 0.15,
has been checked, performing
a fine tuning with different values 0.1 $\leq\,F \,\leq$
0.5 (Figure \ref{figP3}),
both for the rates calculated from Model III (cumulative model)
and Model II (binary massive stars model).
It is evident that $F$ = 0.15 represents quite a reasonable choice
and it is in good agreement with the observations, especially for
the cumulative model (Mod III). This is consistent with the results
by Calura \& Matteucci (2006).

\begin{figure*}
\centering
\includegraphics[height=19pc,width=19pc,bb=28 144 580 700,clip]{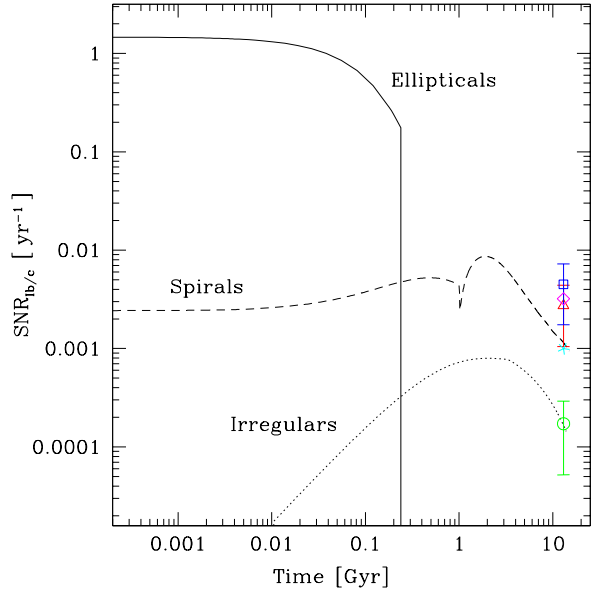} 
\caption{Model III SN rate as a function of time
for galaxies of different morphological type: Ellipticals ({\it solid line}),
Spirals ({\it dashed line}) and Irregulars ({\it dotted line}). $F$=0.15
is assumed.
Data from Cappellaro et al. (1999)
({\it open triangle}), Mannucci et al. (2005) 
({\it open square}), Della Valle (2005) ({\it open rhombus})
and Podsiadlowski et al. (2004) ({\it star});
experimental data for the local SN rate in irregular galaxies 
({\it open circle}) are also shown.}
\label{figP4}
\end{figure*}


Extending the analysis also to elliptical and irregular galaxies,
the results for the cumulative SN rate model are shown in 
Figure \ref{figP4}. A Salpeter IMF has been adopted for
modeling the SN rate in ellipticals and irregulars. $F \,=\,0.15$ is 
assumed. 
The observed rates are indicated
for comparison in the correct units. In the case of irregular galaxies, 
the conversion from SNus is calculated by applying
the B-band luminosity of the main body of the Small Magellanic Cloud (Vangioni-Flam et al. 1980), 
i.e. L$_{B}^{irreg}\,\simeq\,7.8\,\times\,10^{8}\,{\rm L}_{{\rm B}\,\odot}$. \\
Therefore, the rate measured in irregulars by \cite{CAP99}: 
\begin{eqnarray}
{\rm SNR_{Ib/c}}(\,t\sim\,13\,{\rm Gyr}\,)\,=\,\left(0.22\,\pm\,0.16\right)\;{\rm SNu}\, , \nonumber
\end{eqnarray}
is converted in to a value of: 
\begin{eqnarray}
{\rm SNR_{Ib/c}}(\,t\sim\,13\,{\rm Gyr}\,)\,=\,\left(1.72\,\pm\,1.2\right)\,\times\,10^{-4}\;{\rm yr}^{-1}\, . \nonumber
\end{eqnarray}
>From Figure \ref{figP4} it is evident that the SN rate model for an irregular galaxy well
reproduces this observed rate.

\begin{figure*}
\centering
\includegraphics[width=0.95\columnwidth,bb=28 144 580 700,clip]{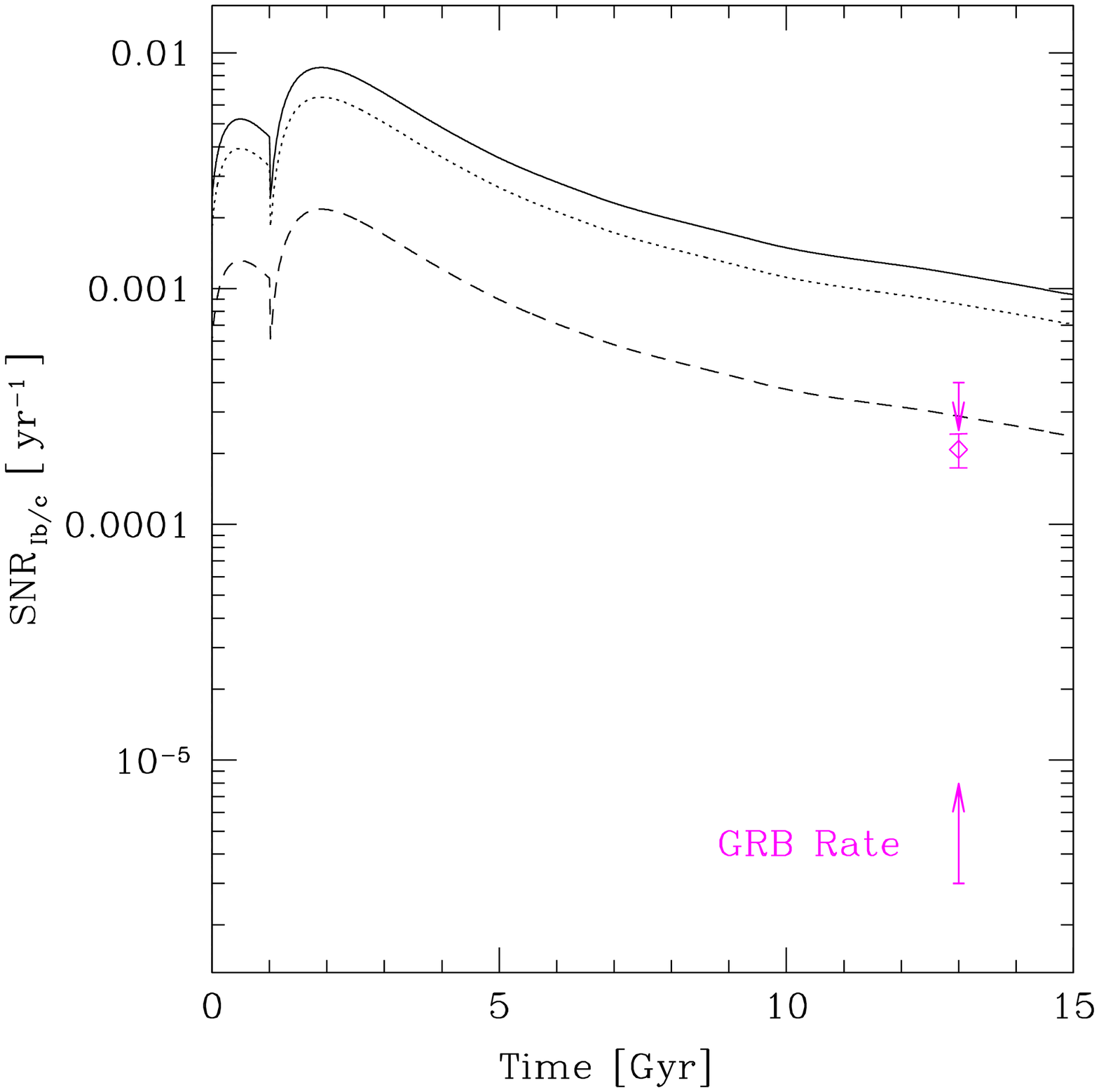} 
\caption{SN rate for a spiral galaxy as a function of time
calculated according to Model I (equation \ref{eqModI}, {\it dotted line});
Model II (equation \ref{eqModII}, {\it dashed line}); 
Model III (equation \ref{eqModIII}, {\it solid line}).
A Scalo 1986 IMF and F = 0.15 are assumed. 
Upper and lower limits for the observed local GRB rates 
(equation \ref{eqGRBminmax}) are
over-plotted ({\it upper and lower arrows}). The {\it open rhombus}
represents a ``typical'' value calculated from a local rate by
\cite{ZHA04} and a beaming correction by uncorrected value by \cite{FRA01} 
(see equation \ref{eqGRBtyp}).}
\label{figP5}
\end{figure*}


\subsection{The local GRB rate}\label{secLGR}
The rate of observed GRBs in a galaxy such as the Milky Way can be
established from the BATSE monitoring as R$_{\rm obs}$.
Sokolov (2001) calculated this value assuming that all or at 
least the main part of long GRBs were associated
with SNe Ib/c. From the observations of known GRB host galaxies 
between 1997 and 2000 (Bloom et al. 2002), 
he deduced that the absolute magnitude
M$_{HostGal}$ ranged from 22 to 28.5 mag. 
The search for direct GRB-SN associations in nearby galaxies
represents a challenge, because the majority of the
SNe related to GRBs are faint (22--26 mag) and in very distant galaxies with
z $\geq$ 0.4--4.5. 
Sokolov (2001) considered the number of galaxies brighter 
than 26 mag in one square degree of the sky, N$_{gal\,<\,26\,mag}\;\approx\;2\,\times\,10^5$
galaxy deg$^{-2}$ (Casertano et al. 2000), and the number of GRBs
observed by BATSE, n$_{\rm GRB}\;\approx\;0.01$ deg$^{-2}$ yr$^{-1}$. 
>From these two quantities he calculated a rate of GRB events 
R$_{\rm obs}\;\sim\,5.0\,\times\,10^{-8}$
galaxy$^{-1}$ yr$^{-1}$. \\
Other local values calculated by different authors
are reported in Table \ref{tabGRB1} ({\it left table}) in units of GRB yr$^{-1}$.

\begin{table*}
\centering
\vspace{4mm}
\caption{}
\begin{tabular}{l | c || l | c}
\hline
\hline
\multicolumn{2}{p{190pt}||}{\centering {\bf Local GRB Rates R$_{\mathbf {obs}}$}} & \multicolumn{2}{p{190pt}}{\centering {\bf GRB Beaming Factor}} \\
\hline
\multicolumn{1}{p{95pt}|}{\bf Author} & \multicolumn{1}{p{95pt}||}{\centering $\mathbf {\left[GRB yr^{-1}\right]}$} & \multicolumn{1}{p{95pt}|}{\bf Author} & \multicolumn{1}{p{95pt}}{\centering $\mathbf {\,\,\left< f_b^{-1}\right>}$} \\
\hline
Sokolov (2001)                 &  $\sim$ 5.0 $\times$ 10$^{-8}$       &   \cite{SCH01} &  $\sim\,100$     \\
Zhang \&\ M\'esz\'aros (2004)  &  $\sim$ 4.0 $\times$ 10$^{-7}$       &   \cite{FRA01} &  $520\,\pm\,85$  \\
Firmani et al. (2004)          &  $\sim$ 2.0 $\times$ 10$^{-7}$       &   \cite{GUE03} &  $75\,\pm\,25$   \\
Della Valle (2005)             &  $\sim$ 3.8 $\times$ 10$^{-7}$       &   \cite{YON05} &  $\sim\,340$     \\
Le \&\ Dermer (2006)           &  $\sim$ 1.5 - 1.9 $\times$ 10$^{-6}$ &   \cite{SOD05} & $13\,\lesssim\,\left< f_{b}^{-1}\right>\,\lesssim\,10^4$ \\
\hline
\hline
\end{tabular}
\label{tabGRB1}
\end{table*}

\subsubsection{``True'' GRB rates}\label{SubSecTGRB}
The so-called true GRB rate is given by:
\begin{eqnarray} 
{\rm R}_{\rm GRB}\,=\,\left< f_b^{-1}\right>\,{\rm R}_{obs}\, .
\end{eqnarray}
where the quantity  $f_b^{-1}$ is the beaming factor (Sari et al. 1999),
which accounts for the fact that a GRB does not light up the full celestial 
sphere but rather a fraction.
Since the possible corrections for the beaming factor
cover a large range of values (see Table \ref{tabGRB1}, {\it right table}), 
only plausible upper and lower limits can be estimated for the local GRB rate:
\begin{eqnarray}\label{eqGRBminmax}
{\rm R}_{\rm GRB,\,min} & \simeq & \left< f_{b,\,{\rm min}}^{-1}\right>\,\cdot\,{\rm R}_{\rm obs,\,min} \nonumber \\
                        & \sim & 3\,\times\,10^{\,-6}\;{\rm yr}^{\,-1} \nonumber \\
{\rm R}_{\rm GRB,\,max} & \simeq & \left< f_{b,\,{\rm max}}^{-1}\right>\,\cdot\,{\rm R}_{\rm obs,\,max} \\
                        & \sim & 4\,\times\,10^{\,-4}\;{\rm yr}^{\,-1}\, , \nonumber
\end{eqnarray}
where only values calculated after 2004 were considered.

A ``typical'' value that is often considered in literature
is given by the local rate calculated adopting the 
uncorrected value by \cite{ZHA04} and the beaming factor by  \cite{FRA01}, namely:
\begin{eqnarray}\label{eqGRBtyp}
{\rm R}_{\rm GRB} & \simeq & \left(\,520\,\pm\,85\,\right)\,\cdot\,\left(\,4.0\,\times\,10^{-7}\,\right)\;{\rm yr}^{-1} \\
                 & \sim & \left(\,2.08\,\pm\,0.34\,\right)\,\cdot\,10^{\,-4}\;{\rm yr}^{\,-1}\, . \nonumber
\end{eqnarray}
Figure \ref{figP5} shows the distributions of the three SN rate Models 
(from equations \ref{eqModI}, \ref{eqModII} and \ref{eqModIII}) 
for a spiral galaxy comparedwith the observed local GRB rates. Again, 
a Scalo (1986) IMF and $F=0.15$ are assumed for Model I--III.
The {\it arrows} mark the lower and upper limits for the
GRB rates as calculated in equations \ref{eqGRBminmax}, 
while the {\it open rhombus} represents 
the ``typical'' GRB rate calculated in equation \ref{eqGRBtyp}.

As one can see from Figure \ref{figP5}, the SN rate Models poorly reproduce the
observed local GRB rates. However, uncertainties regarding
Type Ib/c SN progenitors masses allow us to perform
a tuning of Model I (massive single star model, see equation \ref{eqModI}).
In particular, we considered different mass ranges for the progenitors of 
Type Ib/c SNe, namely: $\Delta_{\rm M1}$ = 25--100 M$_{\odot}$; 
$\Delta_{\rm M2}$ = 40--100 M$_{\odot}$; $\Delta_{\rm M3}$ = 
60--100 M$_{\odot}$; $\Delta_{\rm M4}$ = 80--100 M$_{\odot}$;
$\Delta_{\rm M5}$ = 90--100 M$_{\odot}$. 
The behaviours of the SN rate calculated
with Model I and Model III in spirals are then shown in Figure \ref{figP6} 
({\it left} and {\it right panel}, respectively). 
This is allowed by the fact that 
the limiting mass for the formation of a WR star is still uncertain and 
depends on the
stellar mass loss, which in turn depends on the stellar metallicity.

Taking a closer look to Model I, it is evident that the range of observed GRB 
rates,
in particular the one between the ``typical'' value ({\it open circle})
and the lower limit ({\it bottom arrow}), is better reproduced
if M$_{inf}\,\gtrsim\,40$ M$_{\odot}$
is assumed. This is consistent with the observations of the four
GRBs with a clear SN association (see Introduction), which
predict a progenitor mass as large as 40 M$_{\odot}$ in three out of 
four cases.

A last interesting distribution to study is the SN rate for Model III
in galaxies of different morphological types
together with the observed local GRB rates (Figure \ref{figP7}).
Again, a Salpeter IMF is adopted for ellipticals and irregulars.
Here the integration is performed over the mass range 
$\Delta_{\rm M2}$ = 40--100 M$_{\odot}$ for single massive stars giving rise to WRs. 
The parameter $F=0.15$, as previously discussed.
It is evident that in all cases the SN rate is in good agreement
with the observed local GRB rates. It is worth noting that
in the case of irregular galaxies, the predicted SN rate
falls exactly in the range between the ``typical'' local GRB rate 
and its lower limit, which is consistent with the latest 
observations of GRB-SN hosts (e.g., Conselice et al. 2005; Savaglio et al. 2006) 
and with a recent work by \cite{FRU06}, who showed that the host galaxies 
of GRBs appear to be significantly faint and most of them are irregulars.

\begin{figure*}[p!]
\centering
\begin{tabular}{cc}
\includegraphics[width=0.9\columnwidth,bb=18 144 580 700,clip]{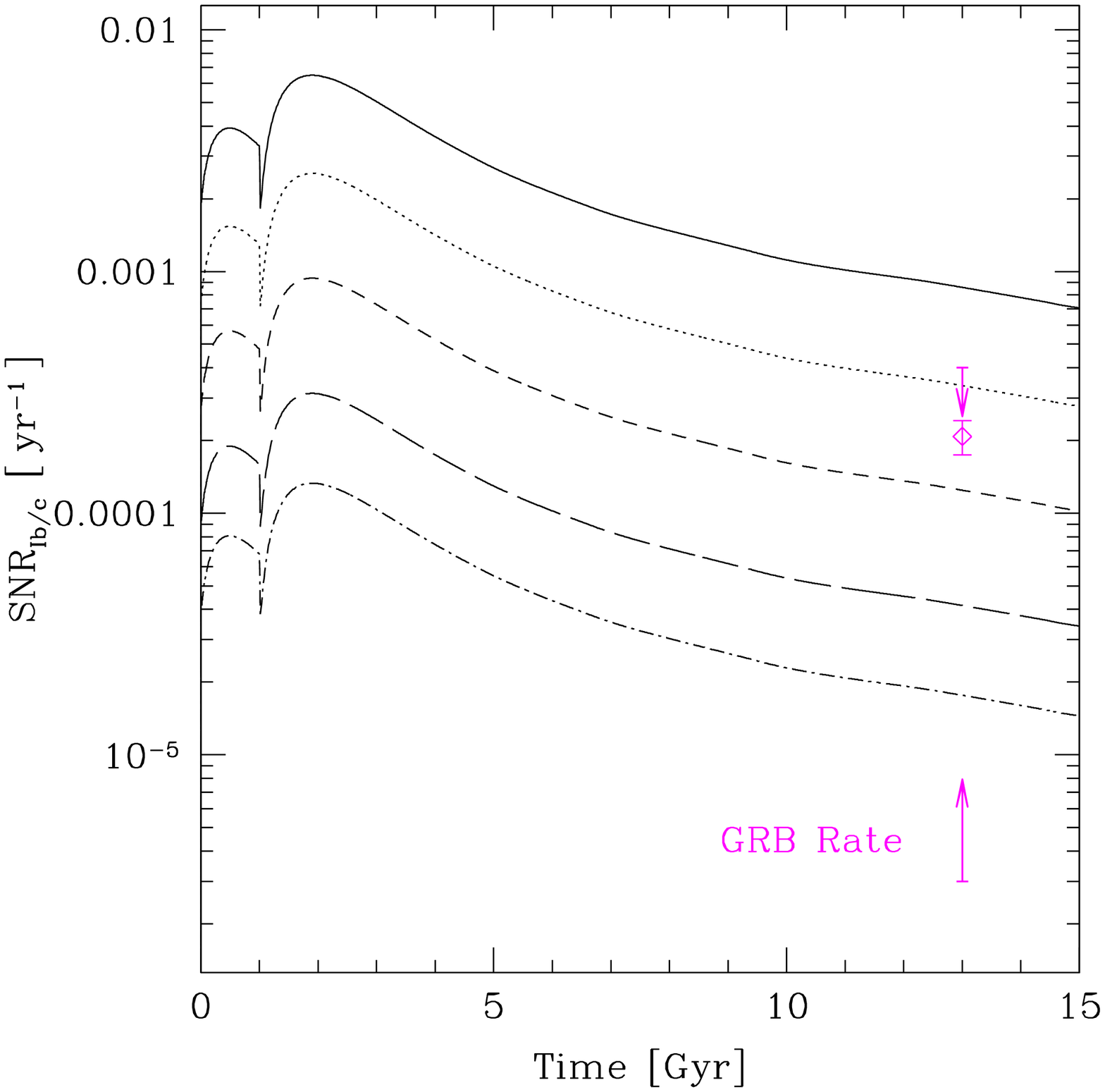}
\includegraphics[width=0.9\columnwidth,bb=18 144 580 700,clip]{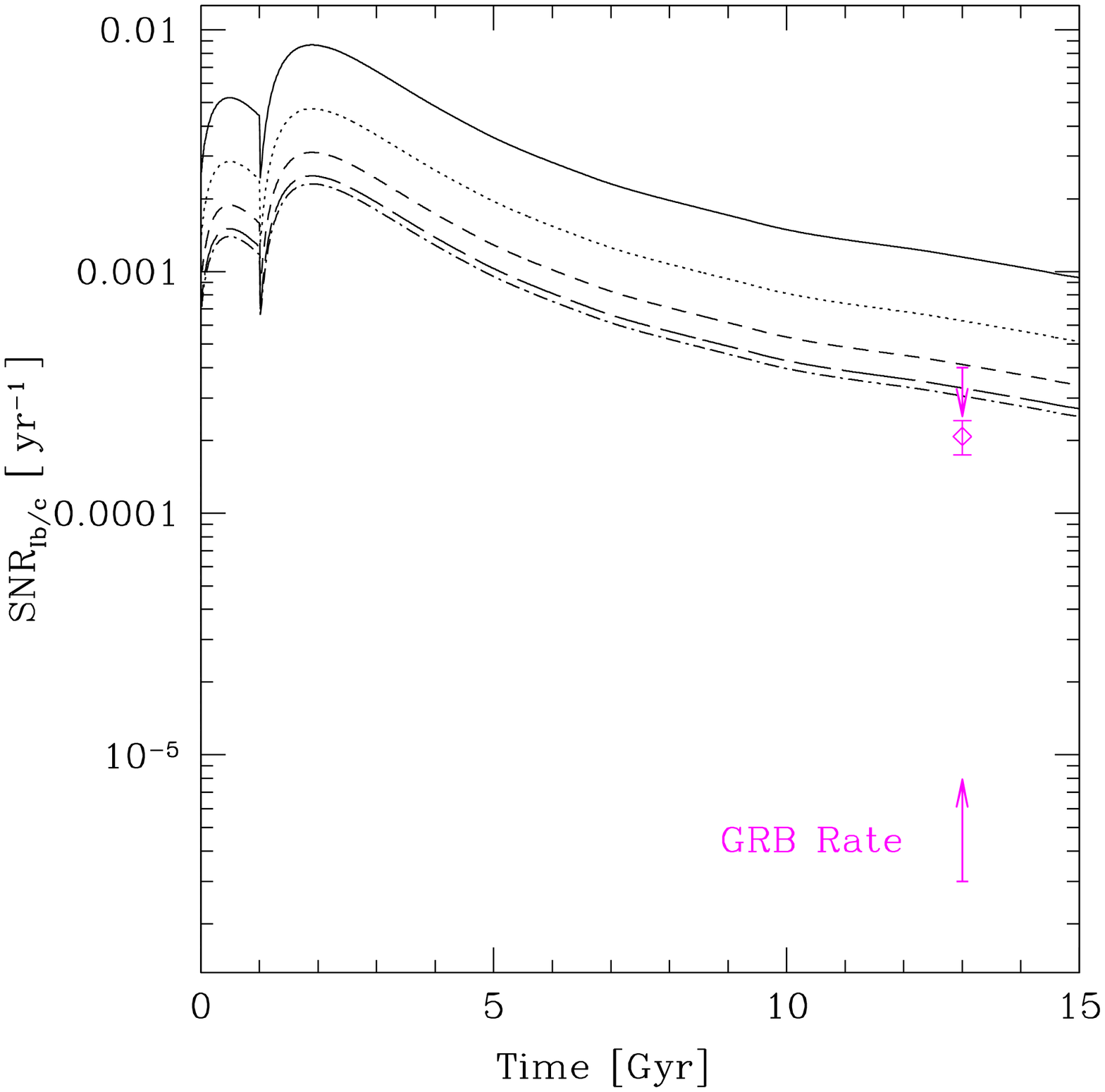} 
\end{tabular}
\caption{SN rate calculated for Model I ({\it Left  Panel}) 
and Model III ({\it Right  Panel}) in a spiral galaxy for five different
mass ranges of existence of Type Ib/c SNe progenitors (single WR stars): 
$\Delta_{\rm M1}$ = 25--100 M$_{\odot}$ ({\it solid line}); 
$\Delta_{\rm M2}$ = 40--100 M$_{\odot}$ ({\it dotted line}); 
$\Delta_{\rm M3}$ = 60--100 M$_{\odot}$ ({\it short-dashed line}); 
$\Delta_{\rm M4}$ = 80--100 M$_{\odot}$ ({\it log-dashed line});
$\Delta_{\rm M5}$ = 90--100 M$_{\odot}$ ({\it dot-dashed line}). 
A Scalo (1986) IMF and a value of $F=0.15$ are assumed.
Upper and lower limits for the observed local GRB rates 
(equation \ref{eqGRBminmax}) are
over-plotted ({\it upper and lower arrows}). The {\it open rhombus}
represents a ``typical'' value calculated from a local rate by
\cite{ZHA04} and a beaming correction by uncorrected value by \cite{FRA01} 
(see equation \ref{eqGRBtyp}).}
\label{figP6}
\vspace{5mm}
\centering
\includegraphics[width=0.98\columnwidth,bb=28 144 580 700,clip]{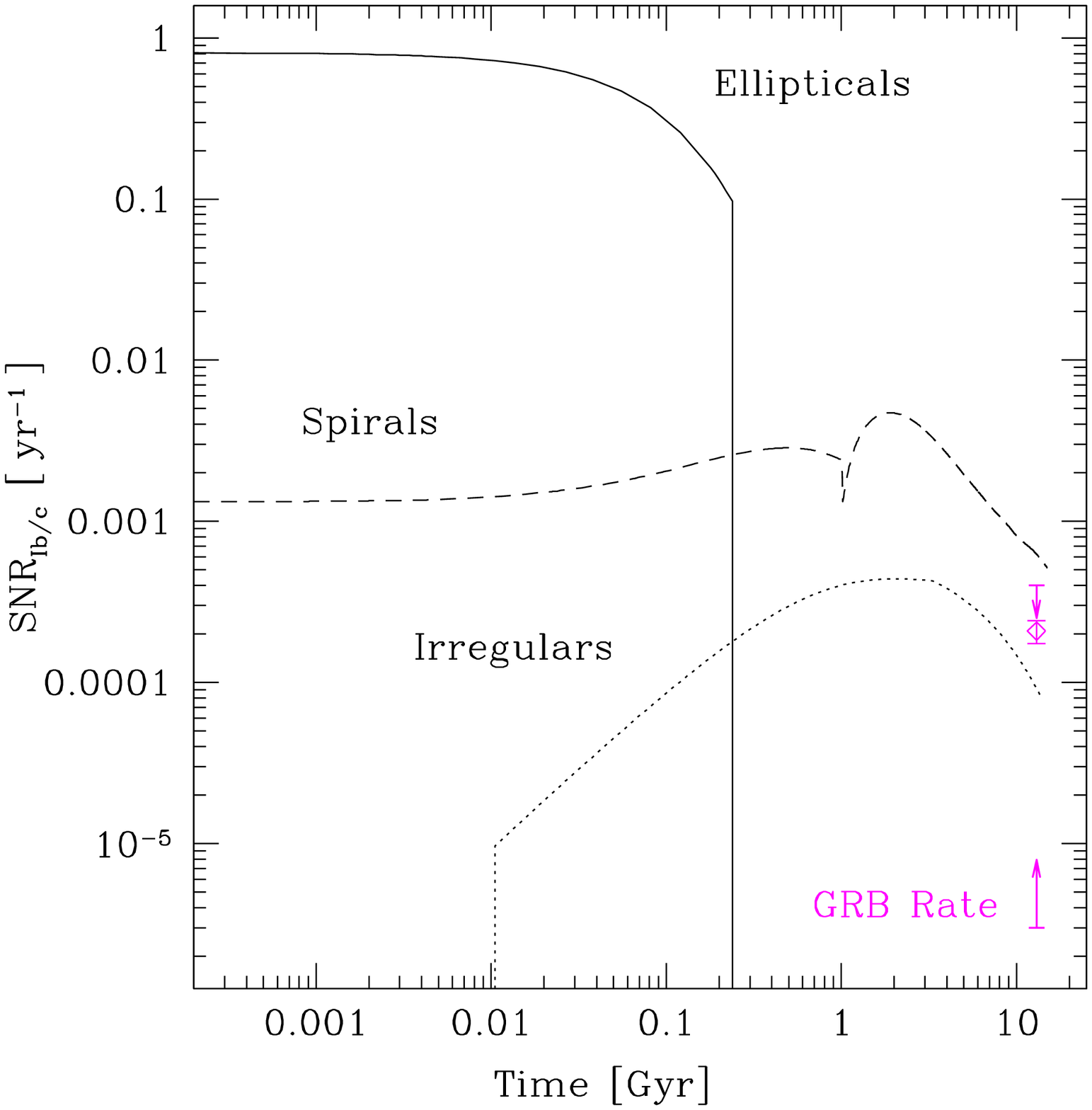} 
\caption{Upper and lower limits for the observed local GRB rates
({\it upper and lower arrows}) compared with Model III SN rate model
for galaxies of different morphological type: ellipticals ({\it solid line}),
irregulars ({\it dotted line}) and spirals ({\it dashed line}).
The {\it open rhombus} represents a ``typical'' value calculated from a local rate by
\cite{ZHA04} and a beaming correction by uncorrected value by \cite{FRA01} 
(see equation \ref{eqGRBtyp}).}
\label{figP7}
\end{figure*}
\begin{table*}
\centering
\caption{Values of the coefficients for $\psi_{\rm MDP_1}(t)$, $\psi_{\rm MDP_2}(t)$
and $\psi_{\rm ST04}(t)$.}
\begin{tabular}{c | c c c c c c c }
\hline
\hline
\multicolumn{1}{p{2cm}||}{\centering {\bf SFR}} & & $\mathbf {a}$ & $\mathbf {b}$ & $\mathbf {c}$ & $\mathbf {d}$ & $\mathbf {e}$ & $\mathbf {d}$ \\
\hline
$\dot{\rho_*}_{\rm MDP_1}(t)$ & & 0.049 & 5 & 0.64 & 0.2 & - & - \\
$\dot{\rho_*}_{\rm MDP_2}(t)$ & & 0.336 & 5 & 0.64 & 0.0074 & 0.0197 & 1.6 \\
$\dot{\rho_*}_{\rm ST04}(t)$ & & 0.182 & 1.260 & 1.865 & 0.071 & - & - \\
\hline
\hline
\end{tabular}
\label{tabPsi}
\end{table*}

\section{Cosmic SN Ib/c and GRB rates and GRB/SNe ratios}
The comoving rate density is a measure of the
number of events occurring per unit comoving volume and time which provides
a census of the number of objects formed at a given redshift and which can
help understanding the object/structure formation in its various
stages of evolution. In the following sections
this quantity will be estimated and discussed for both
Type Ib/c SNe and GRBs.

\subsection{The cosmic SN Ib/c rate}
In order to reproduce cosmic SN Ib/c rate densities, 
the fundamental ingredients are a cosmic SFR density
and a suitable IMF. There are different models which 
can describe the global SFR per unit comoving volume.
In the following, four models will be  reviewed and analysed.

\subsubsection{Cosmic SFR density models}
One model was computed by \cite{CAL03} (hereafter CM03). They
calculated the cosmic SFR density $\dot{\rho}_{\star}$ as a 
function of redshift (Madau's plot)
in the case of a $\Lambda$CDM cosmology and galaxy formation at $z_f$ = 5.
Since the cosmic SFR density is not a directly observable quantity,
CM03 evaluated it by calculating the luminosity density (LD),
$\rho_{\lambda}$ ,
at certain wavelengths and by adopting a universal IMF (the Salpeter one). 

The total LD in a given band is the integrated light
radiated per unit volume from the entire galaxy population
and it needs the luminosity function (LF) to be determined. The LF
represents the distribution of absolute magnitudes for galaxies
of any specified Hubble type and it is often parametrized
according to the form defined by Schechter (1976):
\begin{eqnarray}
\Phi(L)\,\frac{dL}{L_{\star}}\,=\,\Phi_{\star}\left(\frac{L}{L_{\star}}\right)^{-\,\gamma}\,
\exp\left(-\frac{L}{L_{\star}}\right)\,\frac{dL}{L_{\star}}\; ,
\end{eqnarray}
where $\Phi_{\star}$ is a normalization constant related to the number
of luminous galaxies per unit volume, $L_{\star}$ is a characteristic
luminosity and $\gamma$ is the slope of the luminosity function.
The LD stems from the integral over all magnitudes of the observed LF:
\begin{eqnarray}
\rho_L\,=\,\int \Phi(L/L_{\star})\,L/L_{\star}\,d(L/L_{\star}) \, .
\end{eqnarray}

The determination of the SFR density is related to the measures of
star formation in galaxies. CM03 reconstructed the history of cosmic
star formation in the Universe by means of detailed chemical evolution
models for galaxies of different morphological
types, as described before. They normalized the galaxy populations to the B-band LF
observed in the local Universe and studied the redshift evolution
of the LD in various bands (U, B, I and K), calculating galaxy
colors and evolutionary corrections by means of a detailed
synthetic stellar population model (Jimenez et al. 1998).
The cosmic SFR density was computed according to:
\begin{eqnarray}
\dot{\rho}_{\star}(z)\,=\,{\displaystyle \sum_i	} \rho_{B,\,i}(z)\,
\left(\frac{M}{L}\right)_{B,\,i}\,\psi_i(z)\; ,
\end{eqnarray}
where $\rho_{B,\,i}$ represents the B-band LD, 
$\left(\frac{M}{L}\right)_{B,\,i}$ is the
B-band mass-to-light ratio and $\psi_i$ represents the star formation
rate for the galaxies of the {\it i}\-th morphological
type. 

An analytical fit which reproduces the
total cosmic SFR density as a function of time (CM03), is given by:
\begin{eqnarray}
y(t) = \left(0.337 -9.37\cdot t+11.19\cdot t^2\right) \nonumber
\end{eqnarray}
for t $< $ 0.328 Gyr;
\begin{eqnarray}
y(t)  =  \left(-1.57+0.11\cdot t+0.05\cdot t^2\right) \nonumber
\end{eqnarray}
for $0.328< t <2.150$ Gyr; 
\begin{eqnarray}\label{eqYCM}
y(t)  =  \left(-0.836-0.13\cdot t+0.005\cdot t^2\right)
\end{eqnarray}
for $t > 2.150$ Gyr,
where $y(t)= \dot{\rho_{*}}$.

The SFR density for this model is then calculated as:
\begin{eqnarray}\label{eqPsiA}
 \dot{\rho}_{\rm CM03}(t)\,=\,10^{\,y\,(t)}\;{\rm M}_{\odot}\,{\rm yr}^{-1}\,{\rm Mpc}^{-3}\, .
\end{eqnarray}
The CM03 model predicts
a peak at the redshift of galaxy formation due to starbursts in spheroids.
This peak is clearly visible at $z\,\sim\,5$ (Figure 8 of CM03) and it is
followed by a flat behaviour between $z\,\sim\,4.2$ and $z\,\sim\,3$
due to star formation in spiral galaxies. The maximum star formation
in spirals causes a smaller peak at $z\,=\,2$; these galaxies are
also responsible for the decline of the SFR density between 
$z\,=\,2$ and $z\,=\,0$, in agreement with observational data.

Other two cosmic SFR density models considered for
this analysis were calculated following the work done by \cite{MAD98b},
in which the emission history of field galaxies is modeled
at UV, optical and near-IR wavelenghts by tracing the
evolution with cosmic time of their luminosity density,
\begin{eqnarray}
\rho_{\nu}(z)\,=\, \displaystyle \int_0^{\infty}L_{\nu}\,
y_{\nu}(L_{\nu},z)\,dL_{\nu}\, ,
\end{eqnarray}
where $y\,(L_{\nu},z)$ is the best-fitting Schechter luminosity
function in each redshift bin. The integrated light radiated
per unit volume from the entire galaxy population is an average over
cosmic time of the stochastic, possibly short-lived star
formation episodes of individual galaxies, and follows a 
relatively simple dependence on redshift. \cite{MAD98b}  
used a stellar evolution model defined by a time-dependent star
formation rate per unit volume, $\psi(t)$, a universal IMF
and some amount of reddening. In such a system, the luminosity density
at time $t$ is given by the convolution integral:
\begin{eqnarray}
\rho_{\nu}(t)\,=\, p_{\rm esc} \displaystyle \int_0^t l_{\nu}(t')\,
\psi(t-t')\,dt' \, ,
\end{eqnarray}
where $l_{\nu}(t')$ is the specific luminosity radiated per unit
initial mass by a generation of stars with age $l_{\nu}(t')$,
$p_{\rm esc}$ is a time-independent term equal to the
fraction of emitted photons which are not absorbed by dust and
the cosmic SFR density is derived from the observed UV luminosity density.

Based on these considerations, the cosmic SFR model is
given by an analytical fit developed by \cite{MAD98a} (hereafter MDP98):
\begin{eqnarray}\label{eqPsiB}
\dot{\rho_*}
_{\rm MDP_1}(t) = a \left[\,t_9^b e^{-t_9/c}+
d \left(1\,-\,e^{-t_9/c}\right) \right]
\end{eqnarray}
expressed in $M_{\odot}\,yr^{-1}\,Mpc^{-3}$ and  
where $t_9$ is the Hubble time in Gyr,
and the values for the coefficients are given in Table \ref{tabPsi}.

This cosmic SFR was adopted by MDP98
in order to predict the evolution of the
observed comoving luminosity density in the `hierarchical 
clustering' scenario, in which elliptical galaxies form
continuously from the merger of disk-bulge systems or other ellipticals
and most galaxies never experience star formation rates in excess of a few
solar masses per year. 

\begin{figure*}[t!]
\centering
\includegraphics[height=19pc,width=19pc,bb=28 144 580 700,clip]{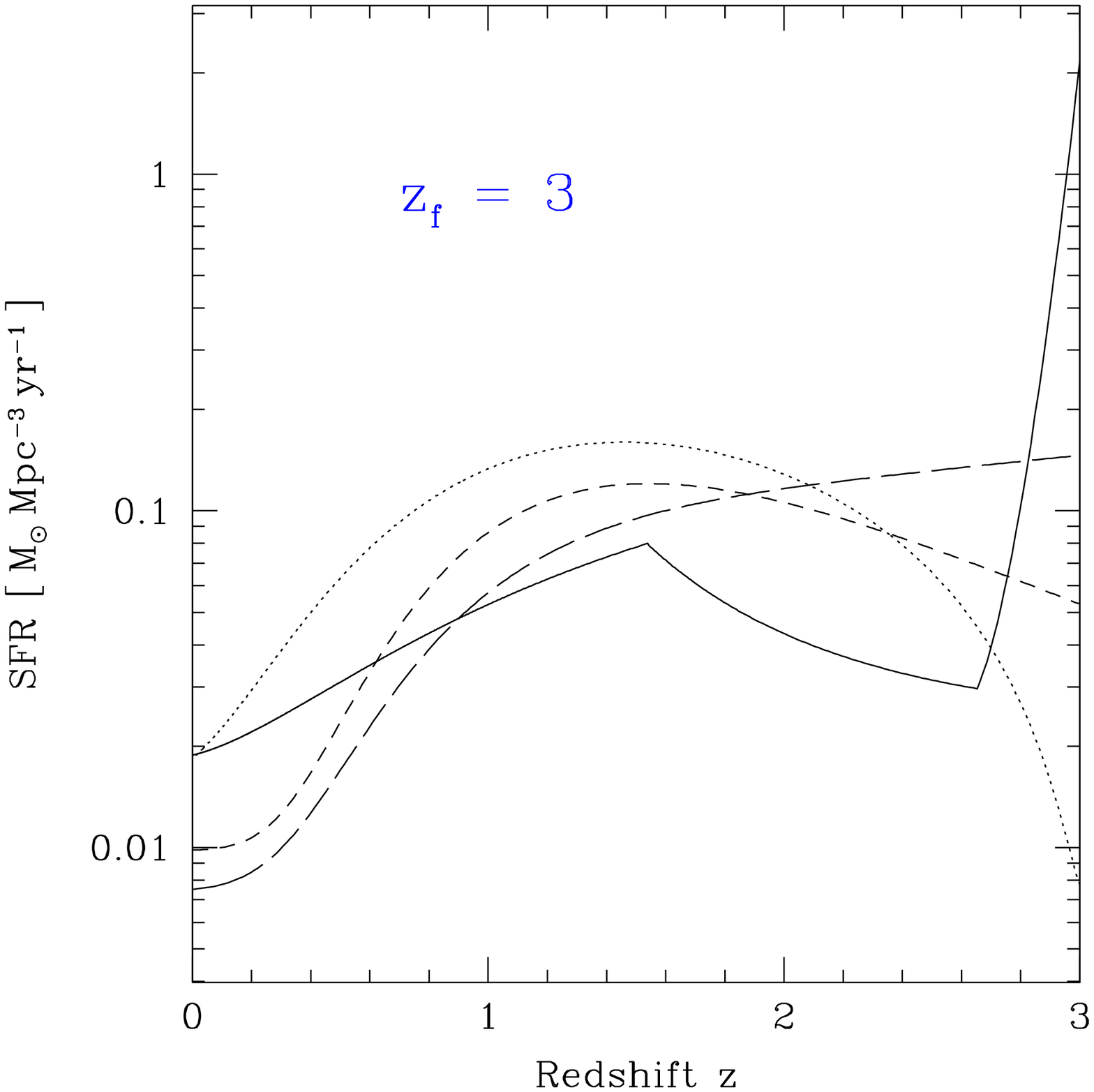} 
\includegraphics[height=19pc,width=19pc,bb=28 144 580 700,clip]{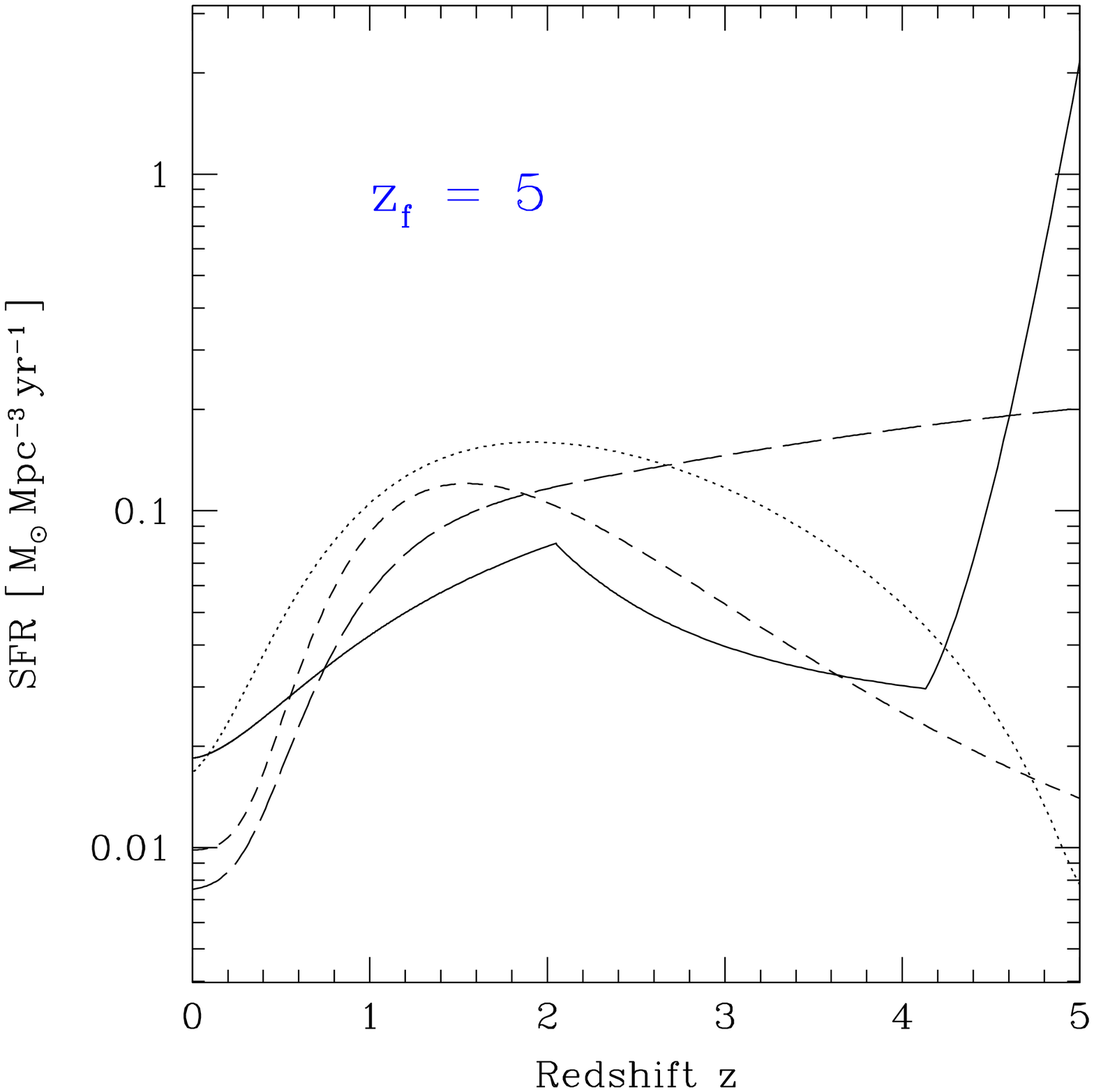} 
\caption{Cosmic SFR densities as a function
of redshift calculated for different epochs of
galaxy formation, namely $z_f=3$ ({\it left panel}) and $z_f=5$
({\it right panel}); 
Model CM03 (equation \ref{eqPsiA}): {\it solid line}; 
Model MDP$_1$ (equation \ref{eqPsiB}): {\it short-dashed line};
Model MDP$_2$ (equation \ref{eqPsiC}): {\it long-dashed line};
Model ST03 (equation \ref{eqPsiD}): {\it dotted line}.}
\label{figSFRcosmoA}
\centering
\includegraphics[height=19pc,width=19pc,bb=28 144 580 700,clip]{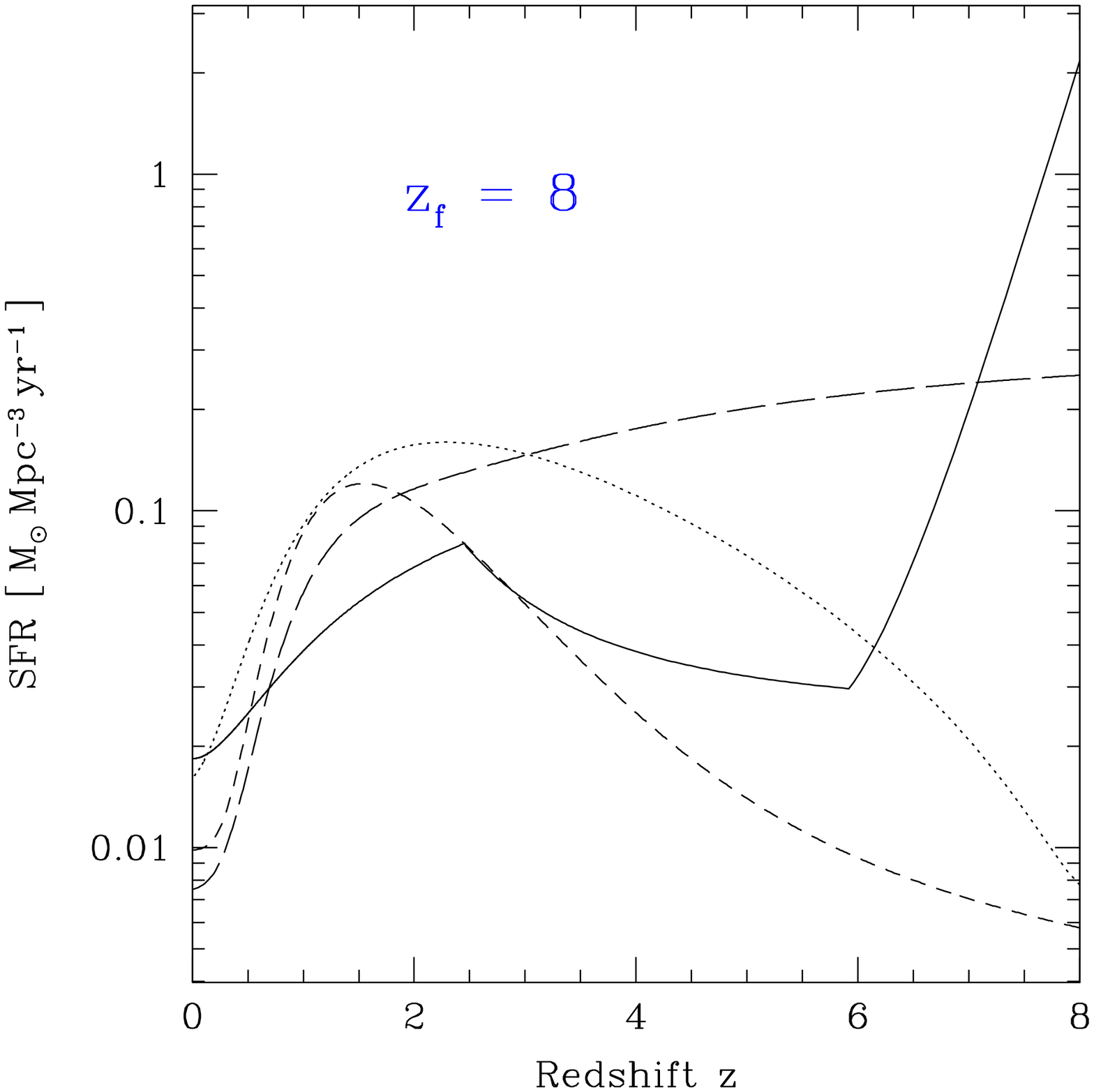} 
\includegraphics[height=19pc,width=19pc,bb=28 144 580 700,clip]{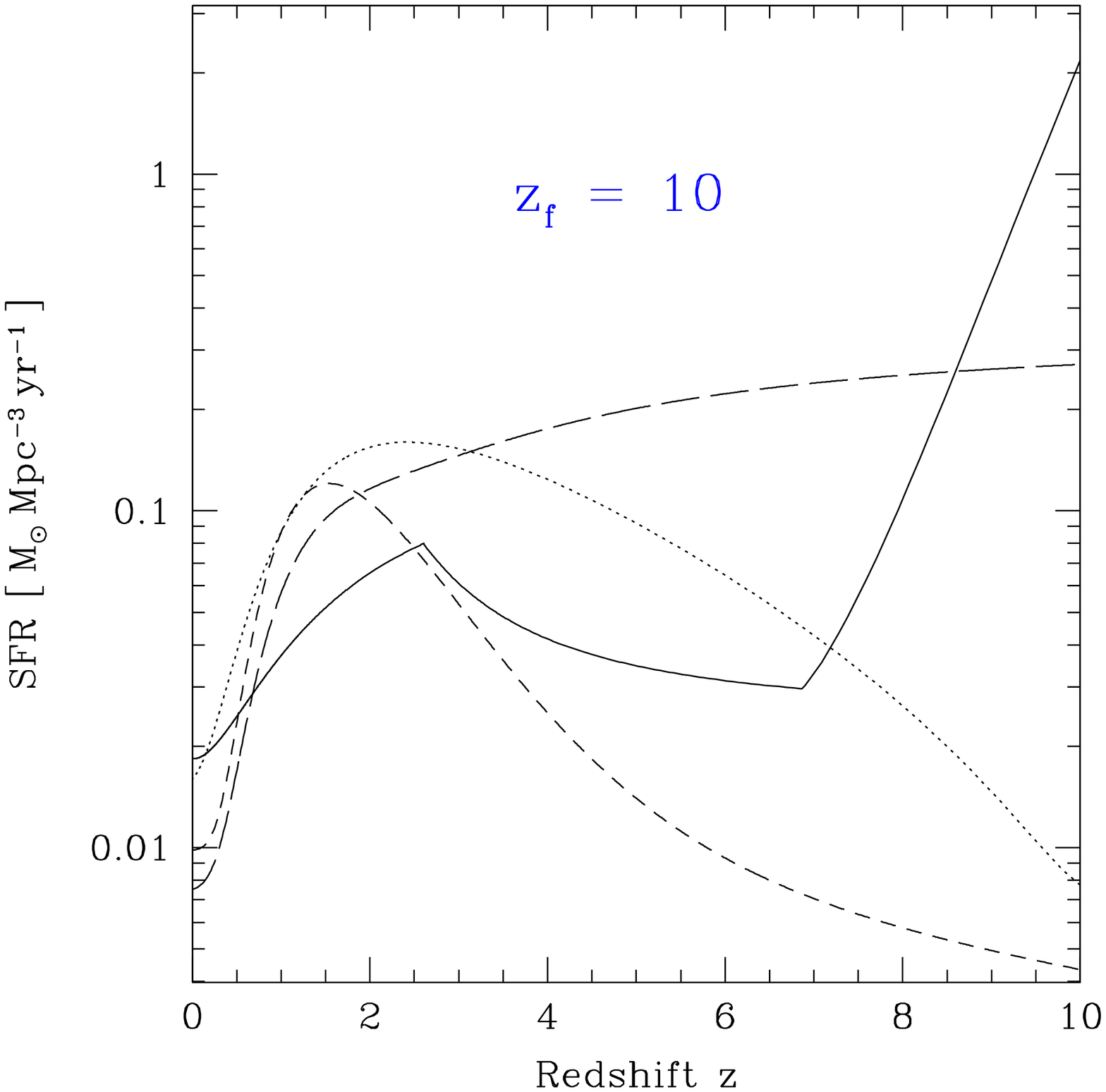} 
\caption{Same as Figure \ref{figSFRcosmoA} calculated for 
different epochs of galaxy formation, namely $z_f=8$ ({\it left panel})
and $z_f=10$ ({\it right panel}). 
Models CM03, MDP$_1$, MDP$_2$, ST04 are indicated as in Figure \ref{figSFRcosmoA}.}
\label{figSFRcosmoB}
\end{figure*}

A second analytical fit by MDP98 consists in a large star formation
density at high redshifts and was designed to mimic the `monolithic
collapse' scenario as in the CM03 model, in which spheroidal systems form early and rapidly,
experiencing a bright starburst phase at high-$z$:
\begin{eqnarray}\label{eqPsiC}
\dot{\rho_*}_{\rm MDP_2}(t)=a\,e^{-t_9/f}+d\left(1-e^{-t_9/c}\right)
+e\,t_9^b\,e^{-t_9/c}
\end{eqnarray}
expressed in units of ${\rm M_{\odot}\,yr^{-1}\,Mpc^{-3}}$,
where the values of the coefficients are given in Table \ref{tabPsi}.

The last model which will be considered in the following analysis,
was computed by Stolger et al. (2004, hereafter ST04),
assuming a modified version of the parametric form of
${\psi}(t)$ as suggested by Madau et al. (1998).
An analytical expression for this model is given by:
\begin{eqnarray}\label{eqPsiD}
\dot{\rho_*}_{\rm ST04}(t)\,=\,a\,\left(t^b\,e^{-t/c}\,+\,d\,
e^{(t\,-\,t_0)/c}\right)
\end{eqnarray}
always in units of ${\rm M}_{\odot}\,{\rm yr}^{-1}\,{\rm Mpc}^{-3}$, 
where $t$ is in Gyr. By fitting the measurements of SFR from
several surveys \cite{GIA04}, ST determined the coefficients of
the function as summarized in Table \ref{tabPsi}.
This model takes the corrections for
extinction into account. Here $t$ is the age of 
the Universe and $t_0$ = 13.47 Gyr corresponds to $z\,=\,0$.

All the already described models are plotted together as functions
of redshift in Figures \ref{figSFRcosmoA} and \ref{figSFRcosmoB} assuming 
four different epochs of galaxy formation $z_f$, namely 3, 5, 8 and 10. 
The different behaviours
of the cosmic SFR following the `monolithic collapse' scenario 
(Models CM03 and MDP$_2$)
or the `hierarchical clustering' scenario (Model MDP$_1$ and ST04) are
clearly visible toward higher redshifts.

\begin{figure*}[t!]
\centering
\includegraphics[width=0.95\columnwidth,bb=18 144 580 700,clip]{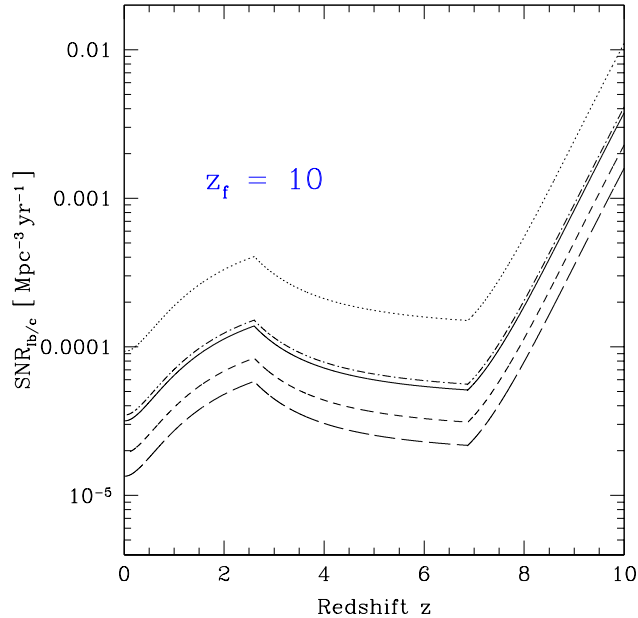} 
\caption{Cosmic SN rate for Model III based on the cosmic SFR
by CM03 and different IMFs: 
Salpeter (1955) (equation \ref{eqSalp}, {\it solid line}); 
``top-heavy'' (equation \ref{eqTop}, {\it dotted line});
Tinsley (1980) ({\it short-dashed line});
Scalo (1986) (equation \ref{eqSca86}, {\it long-dashed line});
Scalo (1998) ({\it dot-dashed line}).
Redshift of galaxy formation: $z_f\,=\,10$.}
\label{figSNR10}
\end{figure*}

\subsubsection{Cosmic SN$_{\mathbf {Ib/c}}$ models}
The strategy for estimating the cosmic SN rate is the same as the one
adopted for the local case in Section \ref{SubSecSNR}. 
Equations \ref{eqModI}, \ref{eqModII} and \ref{eqModIII}
for Model I--III, respectively, can be calculated considering 
different cosmic SFRs, IMFs and various mass ranges.
>From now on, the analysis will focus on the cosmic SN rate calculated for
Model III, i.e. taking both 
massive single stars and massive binary stars into account, 
with a SFR by CM03. This is in fact the Model which gives the best fit to 
the assumed Type Ib/c SN rates in galaxies (Section 1.1).
Moreover, the SFR will be computed assuming
galaxy formation taking place at $z_f\,=\,10$.

This choice of $z_f$ is motivated by the observation
of the most distant quasars (Staguhn et al. 2005), suggesting
massive starbursts up to $z \sim 5$, together with an interesting result
by Mobasher et al. (2005).
Searching for high-redshift $J$--band
``dropout'' galaxies in the portion of the Great 
Observatories Origins Deep Survey
(GOODS) southern field that is covered
by extremely deep imaging from the Hubble Ultradeep Field 
(HUDF), they found evidence for a massive galaxy, HUDF-JD2,
whose spectral energy distribution shows distinctive features which are
consistent with those of a galaxy at $z \sim 6.5$, observed several
hundred million years after a powerful burst of star formation.
The best-fitting models adopted by Mobasher et al. 
placed the formation of the bulk of the stars at $z > 9$.  
For completeness, they also reported alternative models of dusty
galaxies at $z \approx 2.5$. However, these models provided significantly
poorer fits to the photometric data.  
Mobasher et al. suggested that, if the high-redshift
interpretation is correct, HUDF-JD2 is an example of a galaxy that
formed by a process strongly resembling traditional models of
`monolithic collapse'.

An interesting test is to model the SN rate for
different IMFs. Figure \ref{figSNR10} shows its
behaviour as a function of the redshift, for various 
choices of the IMF, namely
a Salpeter IMF; 
a ``top-heavy'' IMF;
a Tinsley (1980) IMF;
a Scalo (1986) IMF, and
a Scalo (1998) IMF.
The Salpeter IMF, which lies at an intermediate position
among the different models, is chosen as the best IMF
for performing further analysis.

\subsection{The cosmic GRB rate}\label{secCGR}

The effort to find a redshift estimate for bursts without 
optical afterglows represents an open research 
branch in the GRB cosmology field. 
The redshift distribution, together with the LF, can
provide important insights not only into the physics of
the individual objects themselves, but also into the evolution of
the matter in the Universe. 

Often, when doing large statistical studies of a
particular class of objects, the LF and redshift
distribution are assumed to be independent quantities; that is, the
luminosity function of the sources are assumed to be the same for all
redshifts. This makes the analysis easier when one has limited
information (e.g. significant data selection effects) obscuring a
direct interpretation of the measured distributions of luminosity and
redshift.  However, it has been shown that this assumption is not
valid for many astrophysical objects, i.e. quasars (Boyle 1993),
as well as for GRBs.

By studying the prompt and especially the afterglow emission
in great detail, the GRB community realized from the very first bursts
with measured redshifts that the GRB LF was {\it not} very
narrow and in fact it exhibited a rather large dispersion,
hence preventing the flux of a GRB to be used as a standard candle
from which to infer a redshift for the source.

In the last years, various types of ``standard-candle relationships''
for GRBs have been discussed, from which a redshift can be inferred based
on common GRB observables. For example, Lloyd-Ronning, Fryer and Ramirez-Ruiz 
(2002, hereafter LFR02) studied GRB formation rates as the result of a
redshift distribution estimated by Fenimore \&\ Ramirez-Ruiz (2000), by means of
a ``Variability-Luminosity'' correlation.
Using a $\tau$ statistical method (Lynden-Bell 1971, Efron \& Petrosian 1992), LFR02 obtained
a GRB comoving rate density as:
\begin{eqnarray}\label{eqRoLFR}
\rho_{LFR02}(z) \, = \, \rho_0 \,
\left\{
\begin{array}{lrccc}
(\,1+z\,)^{\,\sim\,3} & {\rm for}\quad & z & \lesssim & 2 \\
(\,1+z\,)^{\,\sim\,1} & {\rm for}\quad & z & \gtrsim & 2 \, ,
\end{array}
\right. 
\end{eqnarray}
where $\rho_0$ is the normalization constant.
This distribution was computed in units of Mpc$^{-3}$ yr$^{-1}$ 
with an arbitrary normalization of the curve to $\sim$ 1 at 
$(1+z)\,=\,1$.

Another attempt in the GRB redshift estimate was done by 
Yonetoku et al. (2004, hereafter Y04), adopting a ``Peak Luminosity-Peak
Energy'' correlation. Introducing the same statistical method as LFR02, 
they produced the following GRB rate density:
\begin{eqnarray}\label{eqRoYON}
\rho_{Y04}(z)\,=\,\rho_0\,
\left\{
\begin{array}{lrccc}
(\,1+z\,)^{6.0\,\pm\,1.4} & {\rm for}\quad & z & \lesssim & 1 \\
(\,1+z\,)^{0.6\,\pm\,0.2} & {\rm for}\quad & z & \gtrsim & 1 \, .
\end{array}
\right. 
\end{eqnarray} 

In a recent paper by Matsubayashi et al. (2005, hereafter M05), the 
apparent cosmic GRB rates by LFR02 and Y04 have been studied
and used to derive the true absolute GRB formation rate. The peculiarity
of this work lies in the fact that M05 took into
account the geometrical correction of the jet opening angle,
$\theta_j$, and the jet-luminosity evolution,
finally presenting an analytical formula to calculate the true GRB rate.

M05 adopted the cosmic GRB formation rate proposed by Y04,
performing the normalization of $\rho_{Y04}(z)$ at $z=1$, and
obtained a true comoving GRB rate density given by:
\begin{eqnarray}\label{eqRoMA}
\rho_{M05}(z) & = & \rho_{Y04}(z) \left< f_b^{-1} \right>_z \, .
\end{eqnarray}
They derived the distribution of the beaming factor $\left< f_b^{-1} \right>_z$
in a very detailed way, obtaining:
\begin{equation}
\left< f_b^{-1} \right>_z\,=\,200\,\left( \frac{1+z}{2} \right)^{\alpha-\beta}\, .
\end{equation}
Here $\alpha$ and $\beta$ are the slope indexes of the GRB luminosity evolution
$\lambda(z)$ (see LFR02 and Y04 for a detailed description),
\begin{eqnarray}
\lambda(z)\,\propto\,(1+z)^{\alpha}\, ,
\end{eqnarray}
and the jet-corrected luminosity evolution $L_j(z)$ (Lamb et al. 2005),
\begin{eqnarray}
L_j(z)\,\propto\,(1+z)^{\beta}\, ,
\end{eqnarray}
respectively.
M05 imposed $\left< f_b^{-1} \right>_z\,=\, 200$ at $z=1$. In this
case, the mean jet opening half-angle, defined by 
$\left< \theta_j \right>_z\,\equiv\,\left[\left<f_b^{-1} \right>/2 \right]^{-1/2}$,
is equal to the typical value of 0.1 rad at $z=1$.

The value of $\alpha$ has been studied by many authors. 
LFR02 found $\alpha\,\simeq\,1.4\,\pm\,0.5$, while
Y04 calculated $\alpha\,\simeq\,2.60_{-0.20}^{+0.15}$,
which is quite similar to that of quasars. 
\cite{FIR04} found evidence supporting
an evolving LF where the luminosity scales as
$\alpha\,\simeq\,1.0\,\pm\,0.2$. M05 adopted the value
by Y04, i.e. $\alpha\,=\,2.6$ in his work.

Studying possible values for $\beta$, M05 considered three cases:
a) $\beta=\alpha$ -- no jet-opening angle evolution; 
b) $\beta=\alpha/2$ -- intermediate;
c) $\beta=0$ -- nojet-luminosity evolution.
M05 rejected cases (a) and (c) (see the original work for an
exhaustive description), concluding that intermediate values of $\beta$
in the neighbourhood of $\beta\simeq\alpha/2$, for which both
$L_j$ and $\left< f_b^{-1} \right>$ depend on $z$, may be preferable.
Applying a generalized $\tau$ statistical method, M05 
calculated $\beta_{M05}=2.05_{-0.75}^{+0.53}$. 
Recently, the existence of opening angle evolution of the form
$\propto(1+z)^{\gamma}$ with $\gamma=-0.45_{-0.18}^{+0.20}$ was suggested
by \cite{YON05}. Combining this with the result of $\alpha$ by Y04,
a value of $\beta_{Y04}\,=\,\alpha\,+\,2\gamma\,\simeq\,1.7_{-0.41}^{+0.43}$ 
is obtained.

\subsection{Cosmic SN$_{\mathbf {Ib/c}}$ vs. GRB rates}\label{secConf}
Summing up all the results gathered up to this point,
a comparison between the cosmic SN rate models and the
cosmic GRB rate models is the last important step to make in order
to obtain a complete overview.
As it was explained in the previous Sections,
the cosmic SN rate can be calculated in different
scenarios and for different values of the parameters. Here,
the cosmic SFR by CM03 is considered (`monolithic
collapse scenario'), calculated for a redshift of galaxy formation 
$z_f\,=\,10$,
and in units of M$_{\odot}$ Gpc$^{-3}$ yr$^{-1}$.
Two different IMFs are assumed, namely the Salpeter IMF
and the ``top-heavy'' IMF (see Section \ref{SubSecIMF}). Model III is computed assuming 
the usual mass range $\Delta_{M1}$ = 25--100 M$_{\odot}$ 
and a massive binary parameter $F\,=\,0.15$, 
as previously discussed. 

These SN rate models are compared to the cosmic GRB rate density models 
by LFR02, Y04 and M05. Models by LFR02 and Y04 need to be normalized
to the observed local GRB rate, this time expressed
in units of Gpc$^{-3}$ yr$^{-1}$. A typical value which is found
in the literature is $\rho_0\,=\,\rho\,(z=0)\,=\,0.5$ Gpc$^{-3}$ yr$^{-1}$
(Schmidt 2001). Taking different possible beaming corrections 
into account (see Table \ref{tabGRB1}), a range for the true local GRB rate is obtained:
\begin{eqnarray}\label{eqRoCosmo}
\rho_{0,\,{\rm min}}\, & \sim & \,38\,\pm\,13 \, {\rm Gpc^{-3} \, yr^{-1}} \\
\rho_{0,\,{\rm max}}\, & \sim & \,260\,\pm\,40 \, {\rm Gpc^{-3} \, yr^{-1}} \, . \nonumber
\end{eqnarray}

\begin{figure*}
\centering
\begin{tabular}{cc}
\includegraphics[width=0.95\columnwidth]{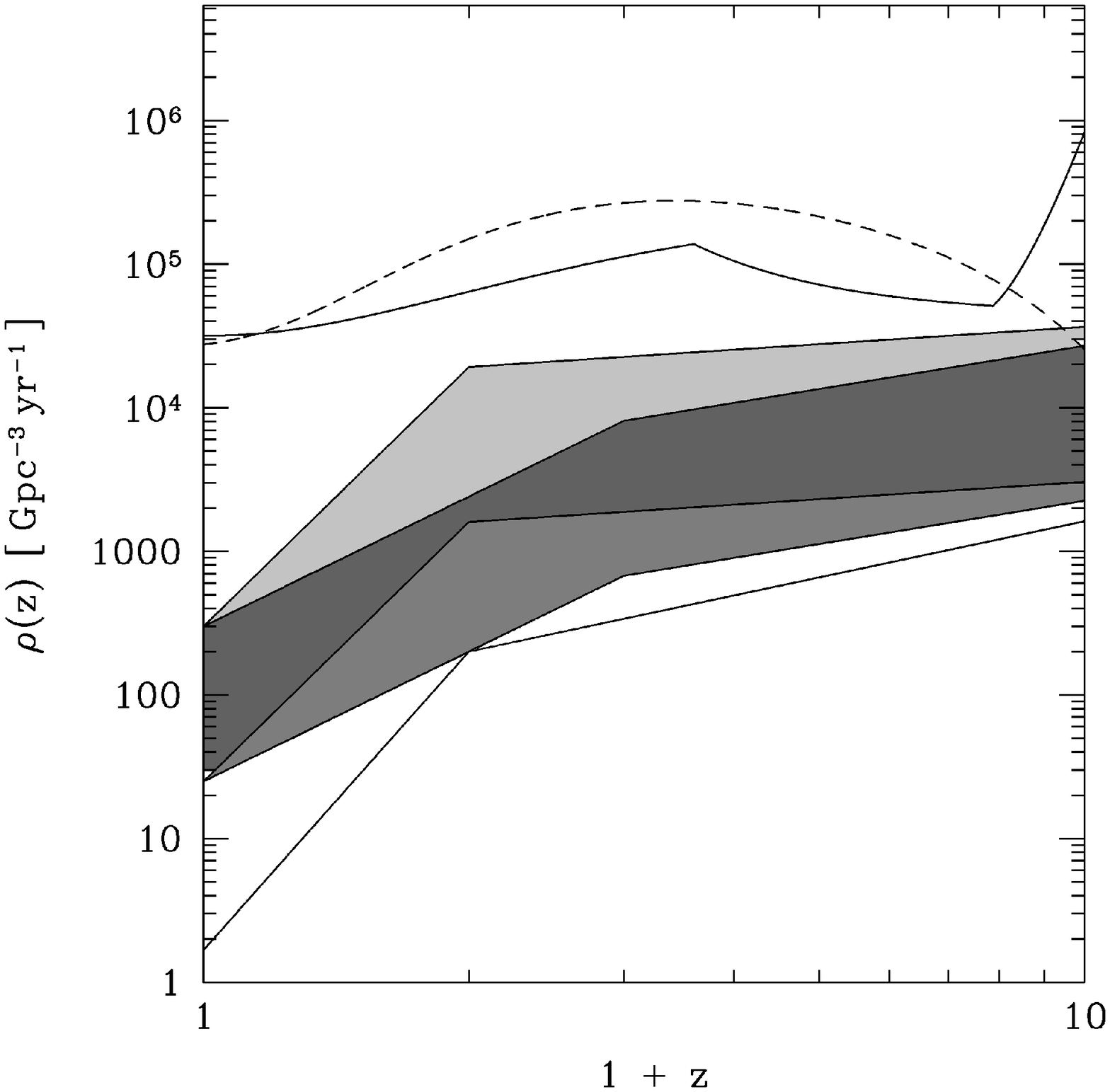}
\includegraphics[width=0.95\columnwidth]{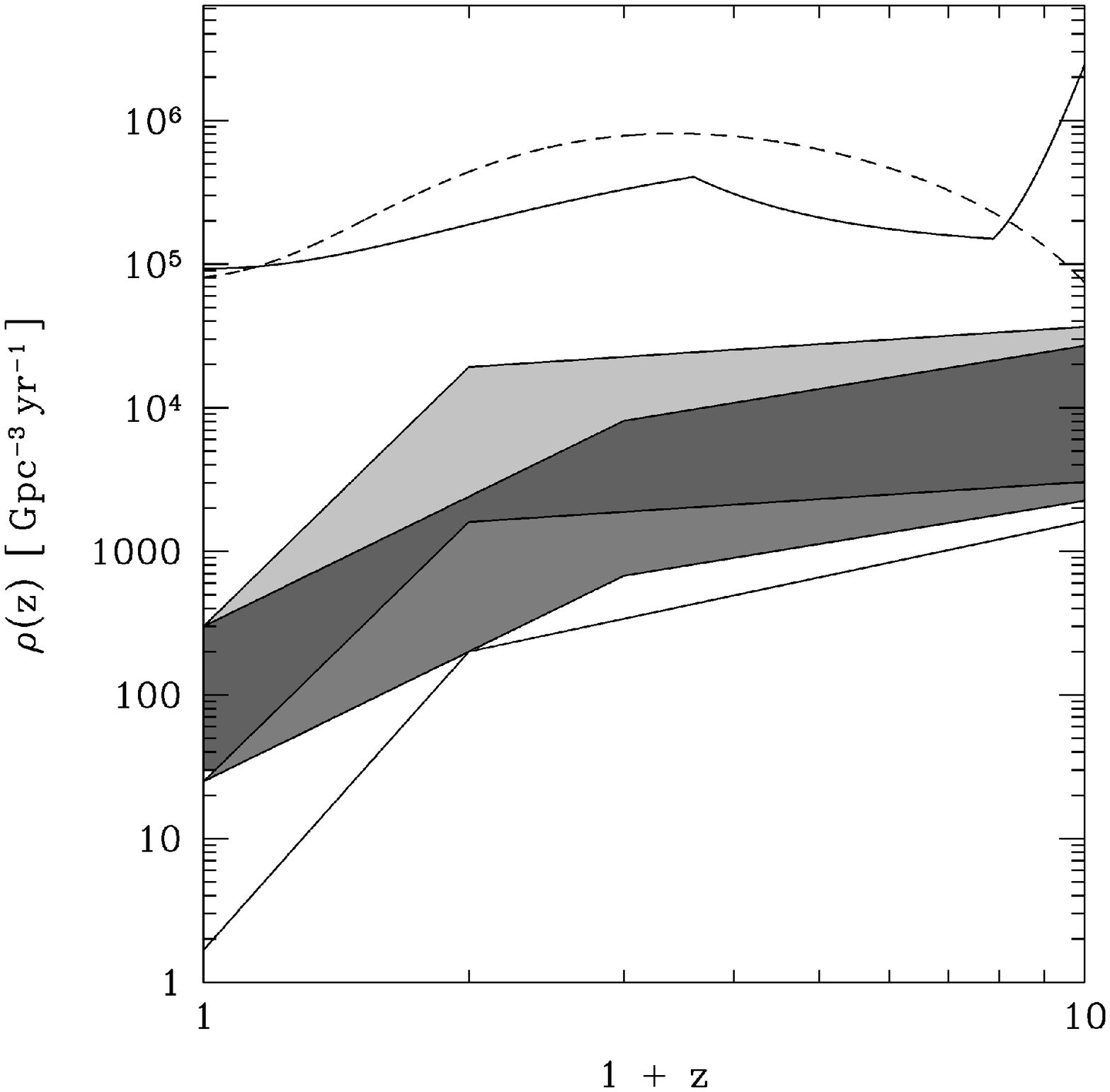}
\end{tabular}
\caption{Absolute cosmic SN rate for Model III calculated for the SFR by CM03 
({\it upper solid curve}) and by ST04 ({\it dashed curve}),
assuming first a Salpeter IMF (equation \ref{eqSalp},
{\it left panel}) and then a ``top-heavy'' IMF (equation \ref{eqTop}, {\it right panel}). 
The plots show also three
absolute comoving GRB rate densities, calculated
according to the models 
by LFR02 ($\rho_{LFR02}(z)$, equation \ref{eqRoLFR}), 
by Y04   ($\rho_{Y04}(z)$, equation \ref{eqRoYON})
and M05 ($\rho_{M05}(z)$, equation \ref{eqRoMA}, {\it lower solid line}). 
Models by LFR02 and Y04 are normalized
in an interval between two local GRB rates, $\rho_{0,\,{\rm min}}$ 
({\it bottom lines}) and $\rho_{0,\,{\rm max}}$
({\it upper lines}, see equation \ref{eqRoCosmo}), covering 
two shaded areas of possible values ({\it light grey, superior } and 
{\it dark grey, inferior areas}, respectively).}
\label{figRo25}
\end{figure*}

Figure \ref{figRo25} shows the behaviours of the cosmic SN rate ({\it upper solid} and
{\it dashed curves}, for models by CM03 and ST04, respectively) 
and the estimated comoving GRB rate densities ({\it shaded areas} for models by LFR02 and
Y04, and {\it lower solid line} for the model by M05). 
The SN rate in the {\it left} and {\it right panels} 
are calculated assuming a Salpeter  and a ``top-heavy'' IMF, respectively.
The GRB models by LFR02 ($\rho_{LFR02}(z)$) and by
Y04 ($\rho_{Y04}(z)$) are normalized in the interval between
$\rho_{0,\,{\rm min}}$ and $\rho_{0,\,{\rm max}}$, corresponding to the ({\it dark grey} 
and {\it light grey areas}, respectively.
The GRB model by M05 ($\rho_{M05}(z)$, {\it lower solidline}) is calculated according to the original
parameters $\alpha\,=\,2.6$, $\beta\,=\,1.7$ and $\rho(0)\,\simeq\,0.01$ Gpc$^{-3}$ yr$^{-1}$.
\begin{table*}
\caption{GRB/SN ratios R$_A$  and R$_B$ (equations \ref{eqRatio}) 
calculated according to the three
models previously discussed (see equations 11, 12 and 14).
{\it TABLE A} displays
Model I--III with the usual choice of the parameters, i.e.
$\Delta_{M_1}$ = 25--100 M$_{\odot}$, $F$ = 0.15
and a Salpeter IMF. Model III is our best model.
{\it TABLE B} focuses on the SN rate
calculated for the massive single star model, Model I,
assuming 4 different mass ranges ($\Delta_{M_2}$--$\Delta_{M_5}$)
and a Salpeter IMF, while {\it TABLE C} highlights
the SN rate calculated for the massive binary star model,
Model II, calculated for 5 different values 
of the binary parameter $F$ and for a Salpeter IMF.
Finally, {\it TABLE D} displays the cumulative model,
Model III, calculated assuming the usual parameters,
$\Delta_{M_1}$ and $F$ = 0.15, but different assumptions
for the IMF.}
\begin{center}
\begin{tabular}{p{4.5cm}|c|c}
\multicolumn{3}{c}{\footnotesize TABLE A} \\
\hline
\hline
 & \multicolumn{1}{p{3.8cm}|}{\centering {$\mathbf{R_{\,A}}$}}%
 & \multicolumn{1}{p{3.8cm}}{\centering {$\mathbf{R_{\,B}}$}} \\
\hline
{\bf Model I}\,\,$^{(\rm a,\,c)}$  & $\sim$ 1.2 $\times$ 10$^{-2}$ & $\sim$ 9.7 $\times$ 10$^{-4}$ \\
{\bf Model II}\,\,$^{(\rm b,\,c)}$ & $\sim$ 4.9 $\times$ 10$^{-2}$ & $\sim$ 4.1 $\times$ 10$^{-3}$ \\
{\bf Model III}\,\,$^{(\rm a,\,b,\,c)}$ & $\sim$ 9.4 $\times$ 10$^{-3}$ & $\sim$ 7.9 $\times$ 10$^{-4}$ \\
\hline
\hline
\multicolumn{3}{l}{\small } \\
\multicolumn{3}{c}{\footnotesize TABLE B} \\
\hline
\hline
{\bf Model I\,\, $^{(\rm c)}$} & \multicolumn{1}{c|}{$\mathbf{R_{\,A}}$}%
 & \multicolumn{1}{c}{$\mathbf{R_{\,B}}$} \\
\hline
$\;$ $\Delta_{\rm M_2}$ = 40--100 M$_{\odot}$ & $\sim$ 2.6 $\times$ 10$^{-2}$ & $\sim$ 2.2 $\times$ 10$^{-3}$ \\
$\;$ $\Delta_{\rm M_3}$ = 60--100 M$_{\odot}$ & $\sim$ 6.5 $\times$ 10$^{-2}$ & $\sim$ 5.4 $\times$ 10$^{-3}$ \\
$\;$ $\Delta_{\rm M_4}$ = 80--100 M$_{\odot}$ & $\sim$ 1.8 $\times$ 10$^{-1}$ & $\sim$ 1.5 $\times$ 10$^{-2}$ \\
$\;$ $\Delta_{\rm M_5}$ = 90--100 M$_{\odot}$ & $\sim$ 4.2 $\times$ 10$^{-1}$ & $\sim$ 3.5 $\times$ 10$^{-2}$ \\
\hline
\hline
\multicolumn{3}{l}{\small } \\
\multicolumn{3}{c}{\footnotesize TABLE C } \\
\hline
\hline
{\bf Model II\,\, $^{(\rm c)}$} & \multicolumn{1}{c|}{$\mathbf{R_{\,A}}$}%
 & \multicolumn{1}{c}{$\mathbf{R_{\,B}}$} \\
\hline
$\;$ $F$ = 0.1 & $\sim$ 7.4 $\times$ 10$^{-2}$ & $\sim$ 6.1 $\times$ 10$^{-3}$ \\
$\;$ $F$ = 0.2 & $\sim$ 3.7 $\times$ 10$^{-2}$ & $\sim$ 3.1 $\times$ 10$^{-3}$ \\
$\;$ $F$ = 0.3 & $\sim$ 2.5 $\times$ 10$^{-2}$ & $\sim$ 2.0 $\times$ 10$^{-3}$ \\
$\;$ $F$ = 0.4 & $\sim$ 1.8 $\times$ 10$^{-2}$ & $\sim$ 1.5 $\times$ 10$^{-3}$ \\
$\;$ $F$ = 0.5 & $\sim$ 1.5 $\times$ 10$^{-2}$ & $\sim$ 1.2 $\times$ 10$^{-3}$ \\
\hline
\hline
\multicolumn{3}{l}{\small } \\
\multicolumn{3}{c}{\footnotesize TABLE D} \\
\hline
\hline
{\bf Model III\,\, $^{(\rm a,\,b)}$} & \multicolumn{1}{c|}{$\mathbf{R_{\,A}}$}%
 & \multicolumn{1}{c}{$\mathbf{R_{\,B}}$} \\
\hline
$\;$ ``top-heavy'' IMF  & $\sim$ 3.2 $\times$ 10$^{-3}$ & $\sim$ 2.7 $\times$ 10$^{-4}$ \\
$\;$ Tinsley IMF        & $\sim$ 1.6 $\times$ 10$^{-2}$ & $\sim$ 1.3 $\times$ 10$^{-3}$ \\
$\;$ Scalo 1986 IMF     & $\sim$ 2.2 $\times$ 10$^{-2}$ & $\sim$ 1.9 $\times$ 10$^{-3}$ \\
$\;$ Scalo 1998 IMF     & $\sim$ 8.6 $\times$ 10$^{-3}$ & $\sim$ 7.2 $\times$ 10$^{-4}$ \\
\hline
\hline
\multicolumn{1}{c}{\it } & \multicolumn{1}{c}{\it } & \multicolumn{1}{c}{\it} \\
\hline
\multicolumn{1}{l}{$^{(a)}$ $\Delta_{\rm M_1}$ = 25--100 M$_{\odot}$} & \multicolumn{1}{c}{\it } & \multicolumn{1}{c}{\it} \\
\multicolumn{1}{l}{$^{(b)}$ $F$ = 0.15} & \multicolumn{1}{c}{\it } & \multicolumn{1}{c}{\it} \\
\multicolumn{1}{l}{$^{(c)}$ Salpeter IMF} & \multicolumn{1}{c}{\it } & \multicolumn{1}{c}{\it}
\end{tabular}
\end{center}
\label{tabSNR/GRB}
\end{table*}
The predicted cosmic SN rates are by far 
larger with respect to the estimated cosmic GRB rate models.
However, this fact is not surprising.
First, local GRB rates suffer from great uncertainties, mainly due to
the beaming corrections.
Moreover, the models applied for the cosmic GRB rate densities
rely upon luminosity indicators and consequent redshift estimates,
which are still far from being confirmed as ``true'' standard-candle
relationships. 
It is worth noting that Figure 11 shows that the cosmic SN Ib/c rate obtained 
by means of the hierarchical scenario (Model ST04) 
predicts a much lower GRB rate at high redshift than the one 
predicted by the monolithic scenario (Model CM03). 

Looking again at Figure \ref{figRo25}, one can see by eye
that this ratio GRB/SNe is almost $\sim\,10^{-2}\,\div\,10^{-3}$, depending
on the choice of the different parameters, which play an important
role in the calculation of the SN rate.
Table \ref{tabSNR/GRB} summarizes in four sub-tables all the different 
analyses which have been
performed in this paper, giving an estimate of the local ratio between
GRB and Type Ib/c SNe according to the various applied models and parameters.

In the tables, two ratios are presented, namely:
\begin{eqnarray}\label{eqRatio}
R_{\,\rm A} & = &  \frac{\rho_{0,\,max}}{SNR_{Ib/c}} ( z = 0 )\\
R_{\,\rm B} & =  & \frac{\rho_{0,\,min}}{SNR_{Ib/c}} ( z = 0 ) \, , \nonumber
\end{eqnarray}
i.e. the ratio are calculated assuming the largest and
the smallest $\left< f_b^{-1}\right>$.
This is to underline the great uncertainty of the local GRB rate  
due to the beaming correction. Still, the results are more
or less consistent with previous works (Podsiadlowski et al. 2004, Della Valle 2005, Le \& Dermer 2006).

While the scatter of possible values for the GRB-SN ratio is
mainly due to uncertainties affecting the observed local GRB rate
and the beaming factor, still the ratio is intrinsically small.
This is motivated by the fact that only a very tiny fraction of 
all massive stars are capable of producing GRBs. 
Type Ib/c SN represent ``obvious'' candidates because they have
lost the H envelope when the collapse of the core occurs, therefore
allowing the ultra-relativistic jets to escape from the massive
progenitor. However this fact does not seem to be sufficient.
According to the SN and GRB rates and to the $\left< f_b^{-1} \right>$
estimates, only 1\% -- 0.1\% 
of SNe$_{Ib/c}$ are able to produce GRBs. 
This fact implies that GRB progenitors must have some other
special characteristic other than being just massive stars. Recently,
several authors have discussed a number of promising mechanisms at
play in the GRB phenomenon, such as the stellar rotation 
(Woosley \& Heger 2006), 
binarity (Podsiadlowski et al. 2004, Mirabel 2004), asymmetric explosions (Maeda et al. 2006)  
and metallicity (Fruchter et al. 2006, Langer \& Norman 2006).

\section{Conclusions}

In this paper we have compared the local GRB rate with local 
Type Ib/c SN rates in galaxies of all morphological type: 
ellipticals, spirals and irregulars. To do that we have adopted 
for each galaxy type a specific history of star formation already 
tested in self-consistent models of galactic chemical evolution, 
which reproduce the main chemical and photometric properties of galaxies 
and in particular the observed SN rates. We have assumed that Type Ib/c SNe, 
which are associated to long-duration, soft GBRs, originate either from 
massive single stars ($M> 25 M_{\odot}$, WR stars) or from massive stars 
in the range $12-20\,M_{\odot}$, which are members of close binary systems.
We made several tests by varying the IMF and the mass range for the existence 
of WR stars. This is justified by the still existing uncertainties 
in the progenitors of WR stars.

Moreover, we computed the cosmic star formation rate, namely the star 
formation rate in a comoving unitary volume of the Universe, by taking 
into account all the galactic types and adopting a 
technique developed by CM03. Then the cosmic star formation rate has 
been used to compute the cosmic Type Ib/c SN rate as a function of 
redshift. In this particular formulation, which is the only one 
distinguishing among different galactic morphological types, 
we assumed that elliptical galaxies, especially the most massive ones, 
form quickly and at high redshift, whereas the spirals and irregulars 
evolve much more slowly. The main driver of galaxy evolution is 
assumed to be the star formation rate, being very fast in ellipticals 
and spheroids in general, and then decreasing going to spirals and 
irregulars. This scenario is different from the classic hierarchical 
scenario of galaxy formation, where ellipticals are assumed to have 
assembled by mergers and preferentially at low redshift. Because of this, 
we have tested also different cosmic star formation rates 
(and Type Ib/c SN rates) including the hierarchical scenario.
We have then derived the GRB rate as a function of redshift 
and compared it with all the cosmic Type Ib/c SN rates  considered.

Our main conclusions can be summarized as follows:

\begin{itemize}
\item The best agreement between the observed and predicted local Type Ib/c 
SN rate in galaxies of different morphological type 
is obtained by assuming that Type Ib/c SNe are both single massive WR stars 
and massive stars in binary systems, in agreement with previous results by 
Calura \&\ Matteucci (2006). Therefore, we conclude that a model including both progenitors
should be preferred to compare with the GRB rate. 

\item When comparing the local GRB rate with the predicted 
local Type Ib/c SN rate, the best agreement is reached under 
the assumptions that WR stars originate in the mass range 
$40-100M_{\odot}$, and the predictions for a typical irregular 
galaxy are those which best fit the local GRB rate. This seems 
to confirm the recent finding by Fruchter et al. (2006) indicating 
that the hosts of GRBs are mainly irregular galaxies. Since 
irregular galaxies are normally metal poor objects, this would 
explain the need of having WR stars only for $M> 40M_{\odot}$ 
as due to the generally less efficient mass loss for metal poor 
stars (see Maeder (1992); Meynet \&\ Mader (2005); see also the most 
recent papers by Yoon \& Langer (2005) and Yoon et al. (2006) 
for the connection between WR stars and GRBs.)  

\item The predicted cosmic SN rate computed with the cosmic 
star formation rate of CM03 follows the derived GRB rate as a 
function of redshift and predicts more GRBs at high redshift 
than hierarchical models. In particular, the predicted rate is 
larger that the cosmic GRB rate, for the whole lifetime of the Universe. 
This is not surprising since 
very likely only a small fraction of Type Ib/c SNe will give rise to GRBs.

\item The ratio between cosmic Type Ib/c SN rate and GRB rate, 
as predicted by our best model, is found to be in the range of 
$\simeq 10^{-2}-10^{-3}$ for a large varieties of cases (different 
IMFs, different assumptions on Type Ib/c SN progenitors). We 
conclude that only $\simeq 0.1-1 \%$ of SNe Ib/c are able to 
produce GRBs. This means that other special characteristics must 
play a role in creating GRBs: stellar rotation, asymmetric 
explosions and metallicity are all possible factors.

\item Because of the high star formation rate in spheroids at 
high redshift assumed in our best models, we predict a higher number 
of GRBs at high redshift than hierarchical models. 
\end{itemize}

\begin{acknowledgements}
E. B. wish to thank Massimo Della Valle for interesting discussions 
and many stimulating comments and an anonymous referee for his/ her 
useful suggestions.  
\end{acknowledgements}

\end{document}